\documentclass{revtex4}

\usepackage{amsthm,amssymb,amsmath}				
\usepackage{graphicx,comment}
\usepackage{float}   
\usepackage{xcolor}

\begin{document}

\title{Schr\"{o}dinger oscillators in a deformed point-like global monopole spacetime and a Wu-Yang magnetic monopole: position-dependent mass correspondence and isospectrality.}
\author{Omar Mustafa}
\email{omar.mustafa@emu.edu.tr}
\affiliation{Department of Physics, Eastern Mediterranean University, G. Magusa, north
Cyprus, Mersin 10 - Turkiye.}

\begin{abstract}
\textbf{Abstract:}\ We show that a specific transformation/deformation
in a point-like global monopole (PGM) spacetime background would yield an
effective position-dependent mass (PDM) Schr\"{o}dinger equation (i.e., a
von Roos PDM Schr\"{o}dinger equation). We discuss PDM Schr\"{o}dinger
oscillators in a PGM spacetime in the presence of a Wu-Yang magnetic
monopole. Within our transformed/deformed global monopole spacetime, we show
that all PDM Schr\"{o}dinger oscillators admit isospectrality and invariance
with the constant mass Schr\"{o}dinger oscillators in the regular global
monopole spacetime in the presence of a Wu-Yang magnetic monopole. The
exclusive dependence of the thermodynamical partition function on the energy
eigenvalues manifestly suggests that the Schr\"{o}dinger oscillators and the
PDM Schr\"{o}dinger oscillators share the same thermodynamical properties as
mandated by their isospectrality. Moreover, we discuss the hard-wall effect
on the energy levels of the PDM Schr\"{o}dinger oscillators in a PGM spacetime
without and with a Wu-Yang magnetic monopole. Drastic energy levels'
shift-ups are observed as a consequence of such hard-wall effect.


\end{abstract}

\maketitle

\section{Introduction}

Various kinds of topological defects depend on the topology of the vacuum
manifold\ and are formed by the phase transition in the early universe \cite%
{Re1,Re2,Re021}. Among such topological defects are the cosmic string \cite%
{Re021,Re3,Re4,Re5,Re6}, domain walls \cite{Re2,Re021}, and global monopole 
\cite{Re7}. Cosmic strings and global monopoles are known to be topological
defects that do not introduce gravitational interactions but they rather
modify the geometry of spacetime \cite{Re3,Re6,Re7,Re8}. Global monopoles
are formed as a consequence of spontaneous global $O(3)$ symmetry breakdown
to $U\left( 1\right) $ and are similar to elementary particles (with their
energy mostly concentrated near the monopole core) \cite{Re7}. They are
spherically symmetric topological defects that admit the general static
metric%
\begin{equation}
ds^{2}=-B\left( r\right) \,dt^{2}+A\left( r\right) \,dr^{2}+r^{2}\left(
d\theta ^{2}+\sin ^{2}\theta \,d\varphi ^{2}\right) .  \label{eq1}
\end{equation}%
Barriola and Vilenkin \cite{Re7} have reported that%
\begin{equation}
B\left( r\right) =A\left( r\right) ^{-1}=1-8\pi G\eta ^{2}-\frac{2GM}{r},
\label{eq2}
\end{equation}%
where $M$ is a constant of integration and in flat space $M\sim M_{core}$ ( $%
M_{core}$ is the mass of the monopole core). \ By neglecting the mass term
and rescaling the variables $r$ and $t$ \cite{Re7}, one may rewrite the
global monopole metric as 
\begin{equation}
ds^{2}=-dt^{2}+\frac{1}{\alpha ^{2}}dr^{2}+r^{2}\left( d\theta ^{2}+\sin
^{2}\theta \,d\varphi ^{2}\right) ,  \label{e01}
\end{equation}%
where $0<\alpha ^{2}=1-8\pi G\eta ^{2}\leq 1$, $\alpha $ is a global
monopole parameter that depends on the energy scale $\eta $,  $G$ is the
gravitational constant, and $\alpha=1$ corresponds to flat Minkowski spacetime  \cite{Re7,Re8,Re9,Re10}. Barriola and Vilenkin \cite%
{Re7} have shown that the monopole, effectively, exerts no gravitational
force. The space around and outside the monopole has a solid deficit angle
that deflects all light. This has motivated several studies, amongst are,
vacuum polarization effects in the presence of Wu-Yang\ \cite{Re101}\
magnetic monopole \cite{Re102}, gravitating magnetic monopole \cite{Re103},
Dirac and Klein-Gordon (KG) oscillators \cite{Re11}, Schr\"{o}dinger
oscillators \cite{Re9}, KG particles with a dyon, magnetic flux and scalar
potential \cite{Re8}, bosons in Aharonov-Bohm flux field and a Coulomb
potential \cite{Re131}, Schr\"{o}dinger particles in a Kratzer potential 
\cite{Re132}, Schr\"{o}dinger particles in a Hulth\.{e}n potential \cite%
{Re1321} , and scattering by a monopole \cite{Re133}. In
general, the influence of topological defects in spacetime on the
spectroscopy of quantum mechanical systems (be it through the introduction
of gravitational field interactions or merely a modification of spacetimes)
has been a subject of research attention over the years. In relativistic
quantum mechanics, for example, the harmonic oscillator is studied in the
context of Dirac and Klein-Gordon (KG) \cite%
{Re11,Re12,Re121,Re13,Re14,Re15,Re16,Re17,Re18,Re19,Re20,Re21,Re211,Re22,Re23,Re24,Re25,Re26,Re27}
in different spacetime backgrounds.

On the other hand, the effective position-dependent mass (PDM) Schr\"{o}%
dinger equation introduced by von Roos \cite{Re271} finds it applications in
nuclear physics, nanophysics, semiconductors, etc \cite%
{Re271,Re272,Re273,Re274}. Where, the von Roos PDM kinetic energy operator 
\cite{Re271} (in $\hbar =2m=1$ units) is given by%
\begin{equation}
\hat{T}=-\frac{1}{2}\left[ m\left( x\right) ^{j}\,\partial _{x}\,m\left(
x\right) ^{k}\,\partial _{x}\,m\left( x\right) ^{l}+m\left( x\right)
^{l}\,\partial _{x}\,m\left( x\right) ^{k}\,\partial _{x}\,m\left( x\right)
^{j}\right] ,  \label{e011}
\end{equation}%
with an effective PDM $m\left( x\right) =mf\left( x\right) $, and $m$ is the mass of Schr\"{o}dinger particle. However, the continuity
conditions at the abrupt heterojunction suggest that $j=l$ \cite%
{Re272,Re273,Re274}, where $j,k,l$ are called the ordering ambiguity
parameters that satisfy the von Roos constraint $j+k+l=-1$ \cite{Re271}.
Recently, it has been shown that under some coordinate
deformation/transformation \cite{Re28} the PDM kinetic energy operator
collapses into%
\begin{equation}
\hat{T}=-m\left( x\right) ^{-1/4}\,\partial _{x}\,m\left( x\right)
^{-1/2}\,\partial _{x}\,m\left( x\right) ^{-1/4},  \label{e012}
\end{equation}%
where $j=l=-1/4$ and $k=-1/2$ (known in the literature as
Mustafa-Mazharimousavi's ordering \cite{Re28,Re281,Re282}. Inspired by
Khlevniuk and Tymchyshyn \cite{Re29} observation that a point mass moving
within the curved coordinates/space transforms into position-dependent mass
in Euclidean coordinates/space, we, hereby, introduce a
deformation/transformation of the global monopole spacetime metric (\ref{e01}%
) and show that the corresponding Schr\"{o}dinger equation transforms into a one-dimensional 
von Roos \cite{Re271} PDM-Schr\"{o}dinger equation. We proceed, under such settings, and discuss
the corresponding effects on the spectroscopic structure of the PDM Schr\"{o}%
dinger oscillators., including the Wu-Yang magnetic monopole and a hard-wall effects.

The organization of our manuscript is in order. In section 2, we
show that a deformation/transformation in the global monopole spacetime metric
(\ref{e01}) would yield an effective position-dependent mass (PDM) Schr\"{o}%
dinger equation. We start with Schr\"{o}dinger particles in the background
of a deformed/transformed global monopole spacetime. We then connect our
findings with the von Roos \cite{Re271}\ PDM Schr\"{o}dinger equation. We
discuss PDM Schr\"{o}dinger oscillators in a global monopole spacetime, in
section 3. We consider, in section 4, the PDM Schr\"{o}dinger oscillators in
a global monopole spacetime in the presence of a Wu-Yang magnetic monopole \ 
\cite{Re101}. Within our deformed/transformed global monopole spacetime
recipe, we show that all our PDM Schr\"{o}dinger oscillators admit
isospectrality and invariance with the Schr\"{o}dinger oscillators in the
regular global monopole spacetime in the presence of a Wu-Yang magnetic
monopole. Nevertheless, the exclusive dependence of the thermodynamical partition function
on the energy eigenvalues manifestly suggests that the Schr\"{o}dinger
oscillators and the PDM Schr\"{o}dinger oscillators have the same
thermodynamical properties as mandated by their isospectrality. We,
therefore, report their thermodynamical properties (e.g., \cite%
{Re9,Re30,Re31,Re32,Re33,Re34,Re35}), in section 5. Such properties are, in
fact, shared by both Schr\"{o}dinger oscillators and PDM Schr\"{o}dinger
oscillators in a global monopole spacetime without and with a Wu-Yang
magnetic monopole.\ In section 6, we discuss the hard-wall effect on the
energy levels PDM Schr\"{o}dinger oscillators in a global monopole spacetime
without and with a Wu-Yang magnetic monopole. The hard-wall confinement is studied by Bakke for a Landau-Aharonov-Casher system \cite{Re351} and for Dirac neutral particles \cite{Re352}, by Castro \cite{Re353} for scalar bosons. and by  Vit\'{o}ria and Bakke \cite{Re354} for the rotating effects on the scalar field in spacetime with linear topological defects, to mention a few. Our concluding remarks are given in section 7. To the best of our knowledge, such a study has not been carried out elsewhere.

\section{Schr\"{o}dinger particles in the background of a
deformed/transformed global monopole spacetime}

Let us consider Schr\"{o}dinger particles interacting with a point-like
global monopole (PGM) with a spacetime metric given by (\ref{e01}) and
subjected to a point canonical transformation (PCT) in the form of 
\begin{equation}
r=\int \sqrt{f\left( \rho \right) }d\rho =\sqrt{q\left( \rho \right) }\rho
\Leftrightarrow \sqrt{f\left( \rho \right) }=\sqrt{q\left( \rho \right) }%
\left[ 1+\frac{q^{\prime }\left( \rho \right) }{2q\left( \rho \right) }\rho %
\right] .  \label{e02}
\end{equation}%
Then the PGM metric (\ref{e01}) transforms into 
\begin{equation}
ds^{2}=-dt^{2}+\frac{f\left( \rho \right) }{\alpha ^{2}}d\rho ^{2}+q\left(
\rho \right) \,\rho ^{2}\left[ d\theta ^{2}+\sin ^{2}\theta \,d\varphi ^{2}%
\right] ,  \label{e03}
\end{equation}%
where $q\left( \rho \right) $ and $f\left( \rho \right) $ are
positive-valued scalar multiplier and $q\left( \rho \right) =1\Rightarrow
f\left( \rho \right) =1$ (i.e., constant mass settings) recovers the PGM
metric (\ref{e01}). Consequently, the corresponding deformed/transformed
metric tensor is 
\begin{equation}
g_{ij}=\left( 
\begin{tabular}{ccc}
$\frac{f\left( \rho \right) }{\alpha ^{2}}$ & $0$ & $0\medskip $ \\ 
$0$ & $q\left( \rho \right) \,\rho ^{2}$ & $0\medskip $ \\ 
$0$ & $0$ & $q\left( \rho \right) \,\rho ^{2}\sin ^{2}\theta $%
\end{tabular}%
\right) ;\;i,j=\rho ,\theta ,\varphi ,  \label{e04}
\end{equation}%
to imply 
\begin{equation*}
\det \left( g_{ij}\right) =g=\frac{f\left( \rho \right) }{\alpha ^{2}}%
q\left( \rho \right) ^{2}\,\rho ^{4}\sin ^{2}\theta ,
\end{equation*}%
and 
\begin{equation}
g^{ij}=\left( 
\begin{tabular}{ccc}
$\frac{\alpha ^{2}}{f\left( \rho \right) }$ & $0$ & $0\medskip $ \\ 
$0$ & $\frac{1}{q\left( \rho \right) \,\rho ^{2}}$ & $0\medskip $ \\ 
$0$ & $0$ & $\frac{1\medskip }{q\left( \rho \right) \,\rho ^{2}\sin
^{2}\theta }$%
\end{tabular}%
\right) .  \label{e05}
\end{equation}%
Then, Schr\"{o}dinger equation 
\begin{equation}
\left\{ \left( -\frac{\hbar ^{2}}{2m_{\circ }}\frac{1}{\sqrt{g}}\partial _{i}%
\sqrt{g}g^{ij}\partial _{j}\right) +V\left( \mathbf{\rho },t\right) \right\}
\Psi \left( \mathbf{\rho },t\right) =i\hbar \frac{\partial }{\partial t}\Psi
\left( \mathbf{\rho },t\right) ,  \label{e06}
\end{equation}%
would, with $V\left( \mathbf{\rho },t\right) =V\left( \rho \left( r\right)
\right) $ and $\Psi \left( \mathbf{\rho },t\right) =e^{-iEt/\hbar }\psi
\left( \rho \right) Y_{\ell m}\left( \theta ,\varphi \right) $, yield 
\begin{equation}
\left\{ \frac{\hbar ^{2}}{2m_{\circ }}\left( -\frac{1}{q\left( \rho \right)
\,\sqrt{f\left( \rho \right) }\,\rho ^{2}}\,\partial _{\rho }\left( \frac{%
q\left( \rho \right) \,\,\rho ^{2}}{\sqrt{f\left( \rho \right) }}\,\partial
_{\rho }\right) +\frac{\ell \left( \ell +1\right) }{\alpha ^{2}q\left( \rho
\right) \,\,\rho ^{2}}\right) +\frac{1}{\alpha ^{2}}V\left( \rho \left(
r\right) \right) \right\} \psi \left( \rho \right) =\frac{1}{\alpha ^{2}}%
E\psi \left( \rho \right) ,  \label{e07}
\end{equation}%
where $Y_{\ell m}\left( \theta ,\varphi \right) $ are the spherical
harmonics, $\ell $ is the angular momentum quantum number, and $m$ is the
magnetic quantum number. In a straightforward manner, equation (\ref{e07})
along with our PCT in (\ref{e02}), is transformed into%
\begin{equation}
\left\{ \frac{\hbar ^{2}}{2m_{\circ }}\left( -\frac{1}{r^{2}}\,\partial
_{r}\,r^{2}\partial _{r}+\frac{\tilde{\ell}\left( \tilde{\ell}+1\right) }{%
r^{2}}\right) +\frac{1}{\alpha ^{2}}V\left( r\left( \rho \right) \right)
\right\} \psi \left( r\left( \rho \right) \right) =\mathcal{E}\psi \left(
r\left( \rho \right) \right)  \label{e08}
\end{equation}%
to imply (with $\psi \left( r\right) =R\left( r\right) /r$)%
\begin{equation}
\left[ \frac{\hbar ^{2}}{2m_{\circ }}\left( -\partial _{r}^{2}+\frac{\tilde{%
\ell}\left( \tilde{\ell}+1\right) }{r^{2}}\right) +\frac{1}{\alpha ^{2}}%
V\left( r\left( \rho \right) \right) \right] R\left( r\right) =\mathcal{E}%
R\left( r\right) ,  \label{e09}
\end{equation}%
where $\mathcal{E}=E/\alpha ^{2}$, and%
\begin{equation*}
\tilde{\ell}\left( \tilde{\ell}+1\right) =\frac{\ell \left( \ell +1\right) }{%
\alpha ^{2}}\Longrightarrow \tilde{\ell}=-\frac{1}{2}+\frac{\sqrt{\alpha
^{2}+4\ell \left( \ell +1\right) }}{2\alpha }
\end{equation*}%
(this would retrieve the regular angular momentum quantum number $\ell $ for
a flat Minkowski spacetime at $\alpha =1$). Moreover, the two quantum
mechanical systems in (\ref{e07}) and (\ref{e08}) are isospectral and
invariant. That is, knowing the solution of one of them would immediately
yield the solution of the other. Yet they both share the same energies.

\subsection{Deformed/transformed PGM spacetime metric and position-dependent
mass connection}

Let us use the substitution of%
\begin{equation}
R\left( r\right) =R\left( r\left( \rho \right) \right) =f\left( \rho \right)
^{-1/4}\phi \left( \rho \right)  \label{e09.1}
\end{equation}%
in (\ref{e09}) to obtain, with (\ref{e02}) and $\partial _{r}R\left(
r\right) =f\left( \rho \right) ^{-1/2}\partial _{\rho }\left( f\left( \rho
\right) ^{-1/4}\phi \left( \rho \right) \right) $,%
\begin{equation}
\left\{ -\frac{\hbar ^{2}}{2m}f\left( \rho \right) ^{-1/2}\partial _{\rho
}f\left( \rho \right) ^{-1/2}\partial _{\rho }+\frac{\hbar ^{2}}{2m}\frac{%
\tilde{\ell}\left( \tilde{\ell}+1\right) }{q\left( \rho \right) \,\rho ^{2}}+%
\frac{1}{\alpha ^{2}}V\left( \rho \right) \right\} f\left( \rho \right)
^{-1/4}\phi \left( \rho \right) =\mathcal{E}f\left( \rho \right) ^{-1/4}\phi
\left( \rho \right) .  \label{e09.2}
\end{equation}%
We now multiply this equation, from the left, by $f\left( \rho \right)
^{1/4} $ to obtain%
\begin{equation}
\left\{ -\frac{\hbar ^{2}}{2m}f\left( \rho \right) ^{-1/4}\partial _{\rho
}f\left( \rho \right) ^{-1/2}\partial _{\rho }\,f\left( \rho \right) ^{-1/4}+%
\tilde{V}\left( \rho \right) \right\} \phi \left( \rho \right) =\mathcal{E}%
\,\phi \left( \rho \right) .  \label{e09.3}
\end{equation}%
Where 
\begin{equation}
\tilde{V}\left( \rho \right) =\frac{\hbar ^{2}}{2m}\frac{\tilde{\ell}\left( 
\tilde{\ell}+1\right) }{q\left( \rho \right) \,\rho ^{2}}\,+\frac{1}{\alpha
^{2}}V\left( \rho \right) ,  \label{e09.4}
\end{equation}%
and consequently the effective kinetic energy operator reads%
\begin{equation}
\hat{T}=-\frac{\hbar ^{2}}{2m}f\left( \rho \right) ^{-1/4}\partial _{\rho
}f\left( \rho \right) ^{-1/2}\partial _{\rho }\,f\left( \rho \right) ^{-1/4}.
\label{e09.5}
\end{equation}%
Such kinetic energy operator belongs, with $m\left( \rho \right) =mf\left(
\rho \right) $ (hence the notion of position-dependent mass is, metaphorically speaking, introduced in the process), to the set of von Roos \cite{Re271} PDM kinetic energy operators%
\begin{equation}
\hat{T}_{vR}=-\frac{\hbar ^{2}}{4}\left[ m\left( \rho \right) ^{j}\partial
_{x}\,m\left( \rho \right) ^{k}\,\partial _{x}m\left( \rho \right)
^{l}+m\left( \rho \right) ^{l}\partial _{x}\,m\left( \rho \right)
^{k}\,\partial _{x}m\left( \rho \right) ^{j}\right] ,  \label{e09.6}
\end{equation}%
where $j=l$ ( which is physically acceptable to secure the continuity
conditions at the abrupt heterojunction in condense matter physics) and $%
j+k+l=-1$ ( where, $j,k,l$ are called ordering ambiguity parameters). In
fact, such a point canonical transformation makes the notion \emph{%
"position-dependent mass"} metaphorically unavoidable in the process. On the
other hand, the parametric ordering $j=l=-1/4$ and $k-1/2$ in (\ref{e09.5})
is known in the literature as Mustafa and Mazharimousavi's ordering \cite%
{Re28}. Yet, in a straightforward manner, one may show that the PDM momentum operator \cite{Re281,Re282}%
\begin{equation}
\mathbf{\hat{p}}\left( \mathbf{\rho }\right) =-i\left( \mathbf{\nabla -}%
\frac{\mathbf{\nabla }f\left( \mathbf{\rho }\right) }{4f\left( \mathbf{\rho }%
\right) }\right) \Longleftrightarrow p_{\rho }=-i\left( \partial _{\rho }-%
\frac{f^{\prime }\left( \rho \right) }{4f\left( \rho \right) }\right)
;\;f\left( \mathbf{\rho }\right) =f\left( \rho \right) ,  \label{e09.7}
\end{equation}%
in%
\begin{equation}
\left\{ \frac{1}{2m}\left( \frac{\mathbf{\hat{p}}\left( \mathbf{\rho }%
\right) }{\sqrt{f\left( \mathbf{\rho }\right) }}\right) ^{2}+V\left( \rho
\right) \right\} \phi \left( \rho \right) =\mathcal{E}\,\phi \left( \rho
\right) ,  \label{e09.8}
\end{equation}%
would yield (\ref{e09.3}) with (\ref{e09.4}) in a flat Minkowski spacetime
at $\alpha =1$. Moreover, the two systems (\ref{e09}) and (\ref{e09.3}) are
isospectral as they share the same energy levels and are invariant,
therefore. In what follows we shall use $\hbar =2m=1$ units and discuss some
illustrative examples.

\section{PDM Schr\"{o}dinger oscillators in a global monopole spacetime background}

Let us consider $V\left( r\left( \rho \right) \right) =\omega ^{2}r^{2}$ in (%
\ref{e09}) to obtain%
\begin{equation}
\left[ -\partial _{r}^{2}+\frac{\tilde{\ell}\left( \tilde{\ell}+1\right) }{%
r^{2}}+\tilde{\omega}^{2}r^{2}\right] R\left( r\right) =\mathcal{E}R\left(
r\right) ,  \label{e10}
\end{equation}%
where $\tilde{\omega}=\omega /\alpha $ and $\mathcal{E}=E/\alpha ^{2}$. This
is the radial spherically symmetric Schr\"{o}dinger oscillator equation that
admits exact textbook solution in the form of%
\begin{equation}
R\left( r\right) \sim r^{\tilde{\ell}+1}\exp \left( -\frac{\tilde{\omega}%
r^{2}}{2}\right) \;_{1}F_{1}\left( \frac{\tilde{\ell}}{2}+\frac{3}{4}-\frac{%
\mathcal{E}}{4\tilde{\omega}},\tilde{\ell}+\frac{3}{2},\tilde{\omega}%
r^{2}\right) ,  \label{e101}
\end{equation}%
for the radial part, which is to be finite and square integrable through the
condition that the confluent hypergeometric series is truncated into a
polynomial of order $n_{r}=0,1,2,\cdots $. In this case, 
\begin{equation}
\frac{\tilde{\ell}}{2}+\frac{3}{4}-\frac{\mathcal{E}}{4\tilde{\omega}}%
=-n_{r}\Rightarrow \mathcal{E}=2\tilde{\omega}\left( 2n_{r}+\tilde{\ell}+%
\frac{3}{2}\right) \Rightarrow E=2\alpha \omega \left( 2n_{r}+\frac{\sqrt{%
\alpha ^{2}+4\ell \left( \ell +1\right) }}{2\alpha }+1\right)  \label{e11}
\end{equation}%
for the energies (which is in exact accord with the result reported by Vit\'{o}ria and Belich in Eq. (11) of \cite{Re9}, $\omega =\omega _{VB}/2$).,
and%
\begin{equation}
R\left( r\right) \sim r^{\tilde{\ell}+1}\exp \left( -\frac{\tilde{\omega}%
r^{2}}{2}\right) \,L_{n_{r}}^{\tilde{\ell}+1/2}\left( \frac{\tilde{\omega}%
r^{2}}{2}\right) \Rightarrow \Psi \left( r,\theta ,\varphi \right) =\mathcal{%
N}_{n_{r},\ell }\,r^{\tilde{\ell}}\exp \left( -\frac{\tilde{\omega}r^{2}}{2}%
\right) \,L_{n_{r}}^{\tilde{\ell}+1/2}\left( \frac{\tilde{\omega}r^{2}}{2}%
\right) Y_{\ell m}\left( \theta ,\varphi \right) .  \label{e12}
\end{equation}%
for the radial part of the wave functions, where $\,L_{n_{r}}^{\tilde{\ell}+1/2}\left( 
\tilde{\omega}r^{2}/2\right) $ are the generalized Laguerre polynomials. Hereby, it should be noted that this quantum mechanical system is
isospectral and invariant with the PDM one%
\begin{equation}
\left\{ -f\left( \rho \right) ^{-1/4}\partial _{\rho }f\left( \rho \right)
^{-1/2}\partial _{\rho }\,f\left( \rho \right) ^{-1/4}+\frac{\tilde{\ell}%
\left( \tilde{\ell}+1\right) }{q\left( \rho \right) \,\rho ^{2}}\,+\tilde{%
\omega}^{2}q\left( \rho \right) \,\rho ^{2}\right\} \phi \left( \rho \right)
=\mathcal{E}\,\phi \left( \rho \right) ,  \label{e13}
\end{equation}%
where $f\left( \rho \right) $ and $q\left( \rho \right) $ are correlated through (\ref{e02}), provided that $R\left( r\left( \rho \right) \right) $
is given by (\ref{e09.1}). For example, for a power-low like dimensionless
radial deformation $q\left( \rho \right) =A\rho ^{\sigma }$, we obtain $%
f\left( \rho \right) =A\left( 1+\sigma /2\right) ^{2}\rho ^{\sigma }$; $%
\sigma \neq 0,-2$, and the corresponding PDM Schr\"{o}dinger oscillators system reads%
\begin{equation}
\left\{ -\left( \tilde{A}\rho ^{\sigma }\right) ^{-1/4}\partial _{\rho
}^{-1/2}\left( \tilde{A}\rho ^{\sigma }\right) \,\partial _{\rho }\,\left( 
\tilde{A}\rho ^{\sigma }\right) ^{-1/4}+\frac{\tilde{\ell}\left( \tilde{\ell}%
+1\right) }{A\rho ^{\sigma +2}}\,+\tilde{\omega}^{2}A\rho ^{\sigma
+2}\right\} \phi \left( \rho \right) =\mathcal{E}\,\phi \left( \rho \right),  \label{e14}
\end{equation}%
where $\tilde{A}=A\left( 1+\sigma /2\right)^{2}$.  Such a system represents just one of so many examples on PDM 
Schr\"{o}dinger oscillators interacting with a PGM and share the same eigenvalues, (\ref{e11}) with that in 
(\ref{e10}).  

\section{PDM Schr\"{o}dinger oscillators in a PGM spacetime and
a Wu-Yang magnetic monopole}

In this section, we discuss PDM Schr\"{o}dinger particles a PGM spacetime and a Wu-Yang magnetic monopole. Wu and Yang \cite{Re101} have proposed a magnetic monopole that is free of strings of singularities around it \cite{Re8,Re101,Re102}. They have defined the vector potential $A_{\mu }$
in two regions, $R_{A}$ and $R_{B}$, covering the whole space, outside the
magnet monopole, and overlap in $R_{AB}$ so that%
\begin{equation}
\begin{tabular}{lll}
$R_{A}:0\leq \theta <\frac{\pi }{2}+\delta \medskip ,\;$ & $\;\;r>0,\;\;$ & $%
0\leq \varphi <2\pi ,$ \\ 
$R_{B}:\frac{\pi }{2}-\delta <\theta \leq \pi ,\;$ & $\;\;r>0,\;\;$ & $0\leq
\varphi <2\pi ,$ \\ 
$R_{AB}:\frac{\pi }{2}-\delta <\theta <\frac{\pi }{2}+\delta ,$ & $%
\;\;r>0,\; $ & $0\leq \varphi <2\pi ,$%
\end{tabular}
\label{e15}
\end{equation}%
where $0<\delta \leq \pi /2$. Moreover, the vector potential has a
non-vanishing component in each region given by%
\begin{equation}
\begin{tabular}{ll}
$A_{\varphi ,A}=g\left( 1-\cos \theta \right) ,\;\;$ & $A_{\varphi
,B}=-g\left( 1+\cos \theta \right) $%
\end{tabular}%
,  \label{e16}
\end{equation}%
where $g$ is the Wu-Yang monopole strength and $A_{\varphi ,A}$ and $%
A_{\varphi ,B}$ are correlated by the gauge transformation \cite{Re8,Re102} 
\begin{equation}
A_{\varphi ,A}=A_{\varphi ,B}+\frac{i}{e}S\,\partial _{\varphi
}\,S^{-1}\,;\;S=e^{2iq\varphi },\;q=eg.  \label{e17}
\end{equation}%
We shall, for the sake of simplicity and economy of notation, use the form $%
A_{\varphi }=sg-g\cos \theta $, with $s=1$ for $A_{\varphi ,A}$ and $s=-1$
for $A_{\varphi ,B}$.

Under such settings, equation (\ref{e06}) would now read%
\begin{equation}
\left\{ \left( -\frac{1}{\sqrt{g}}\left( \partial _{i}-ieA_{i}\right) \sqrt{g%
}g^{ij}\left( \partial _{j}-ieA_{j}\right) \right) +V\left( \mathbf{\rho }%
,t\right) \right\} \Psi \left( \mathbf{\rho },t\right) =i\frac{\partial }{%
\partial t}\Psi \left( \mathbf{\rho },t\right) ,  \label{e18}
\end{equation}%
to imply%
\begin{equation}
\left\{ -\frac{\alpha ^{2}}{r^{2}}\,\partial _{r}\left( r^{2}\,\partial
_{r}\right) -\frac{1}{r^{2}}\left( \frac{1}{\sin \theta }\partial _{\theta
}\sin \theta \;\partial \theta +\frac{1}{\sin ^{2}\theta }\left[ \partial
_{\varphi }-ieA_{\varphi }\right] ^{2}\right) +V\left( \mathbf{\rho }\left(
r\right) ,t\right) \right\} \Psi \left( \mathbf{\rho },t\right) =i\frac{%
\partial }{\partial t}\Psi \left( \mathbf{\rho },t\right) ,  \label{e19}
\end{equation}%
where $r$ is given by (\ref{e02}). We may now seek separation of variables for (\ref{e19}) and use the substitution $\Psi \left( \mathbf{\rho },t\right)
=e^{-iEt}\;\psi \left( \rho \right) \;Y_{\tilde{q}\ell m}\left( \theta
,\varphi \right) $,$\;$where $\tilde{q}=sq$ and $Y_{\tilde{q}\ell m}\left(
\theta ,\varphi \right) $ are the Wu-Yang monopole harmonics so that%
\begin{equation}
\left( \frac{1}{\sin \theta }\partial _{\theta }\sin \theta \;\partial
\theta +\frac{1}{\sin ^{2}\theta }\left[ \partial _{\varphi }-ieA_{\varphi }%
\right] ^{2}\right) Y_{\tilde{q}\ell m}\left( \theta ,\varphi \right)
=-\lambda Y_{\tilde{q}\ell m}\left( \theta ,\varphi \right) .  \label{e21}
\end{equation}%
Consequently (\ref{e19}) reduces to%
\begin{equation}
\left\{ -\frac{\alpha ^{2}}{r^{2}}\,\partial _{r}\left( r^{2}\,\partial
_{r}\right) +\frac{\lambda }{r^{2}}+V\left( \mathbf{\rho }\left( r\right)
\right) \right\} \psi \left( \rho \right) =E\psi \left( \rho \right) .
\label{e22}
\end{equation}

At this point, one should first solve for the eigenvalues $\lambda $ of \ (\ref{e21}) using the substitution%
\begin{equation}
Y_{\tilde{q}\ell m}\left( \theta ,\varphi \right) =\exp \left( i\left( m+%
\tilde{q}\right) \varphi \right) \Theta _{\tilde{q}\ell m}\left( \theta
\right) ;\;\tilde{q}=sq=seg,  \label{e20}
\end{equation}%
to obtain%
\begin{equation}
\left( \frac{1}{\sin \theta }\partial _{\theta }\sin \theta \;\partial
\theta -\frac{1}{\sin ^{2}\theta }\left[ m+q\cos \theta \right] ^{2}\right)
\Theta _{\tilde{q}\ell m}\left( \theta \right) =-\lambda \Theta _{\tilde{q}%
\ell m}\left( \theta \right) .  \label{e23}
\end{equation}%
Notably, this equation does not depend on the value of $s$ in $\tilde{q}$ of
(\ref{e20}) (i.e., $\Theta _{\tilde{q}\ell m}\left( \theta \right) =\left[
\Theta _{q\ell m}\left( \theta \right) \right] _{A}=\left[ \Theta _{q\ell
m}\left( \theta \right) \right] _{B}=\Theta _{q\ell m}\left( \theta \right) $
as observed by Wu-Yang \cite{Re101}) and consequently, with $x=\cos \theta $%
, would read%
\begin{equation}
\left\{ \left( 1-x^{2}\right) \,\partial _{x}^{2}-2x\,\partial _{x}-\frac{%
\left( m+q\,x\right) ^{2}}{1-x^{2}}\right\} \Theta _{q\ell m}\left( x\right)
-\lambda \Theta _{q\ell m}\left( x\right) .  \label{e24}
\end{equation}%
Let us define%
\begin{equation}
\Theta _{q\ell m}\left( x\right) =\left( 1-x\right) ^{\sigma/2}\left( 1+x\right)
^{\nu/2}\,P_{q\ell m}\left( x\right) ,  \label{e25}
\end{equation}%
to obtain, with $\sigma=(|m|+q)$ and $\nu=(|m|-q)$ (this choice is motivated by the
fact that the space around a monopole is without singularities and so is the wave function around the monopole \cite{Re101}),%
\begin{equation}
\left( x^{2}-1\right) \,P_{q\ell m}^{^{\prime \prime }}\left( x\right) +
\left[ 2q+2\left( m+1\right) x\right] \,P_{q\ell m}^{^{\prime }}\left(
x\right) +\left( m^{2}+m-q^{2}-\lambda \right) \,P_{q\ell m}\left( x\right)
=0.  \label{e26}
\end{equation}%
The exact solution of which admits the form of hypergeometric functions%
\begin{equation}
P_{q\ell m}\left( x\right) =C\,_{1}F_{1}\left( |m|+\frac{1}{2}\pm \frac{1}{2}%
\sqrt{4q^{2}+4\lambda +1},|m|+1-q,\frac{1}{2}\left( 1+x\right) \right) .
\label{e27}
\end{equation}%
However, to secure finiteness and square integrability of the quantum
mechanical wave functions, we truncate the confluent hypergeometric series
into a polynomial of order $n=0,1,2,\cdots $. In this case, we take%
\begin{equation}
-n=|m|+\frac{1}{2}\pm \frac{1}{2}\sqrt{4q^{2}+4\lambda +1}\Longrightarrow
\lambda =\left( n+|m|\right) \left( n+|m|+1\right) -q^{2}\Longrightarrow
\lambda =\upsilon \left( \upsilon +1\right) -q^{2},  \label{e28}
\end{equation}%
where $\upsilon=n+|m| =\ell =0,1,2,\cdots $ is, without loss of generality, the
angular momentum quantum number. That is, when $q=0$ (i.e., the Wu-Yang
monopole strength $g$ is zero) one should naturally retrieve the eigenvalue
of the regular spherical harmonics as $\lambda =\ell \left( \ell +1\right) $%
. Obviously, this result is in exact accord with that reported by Wu and
Yang \cite{Re8,Re101} who have named $Y_{\tilde{q}\ell m}\left( \theta
,\varphi \right) $ as the monopole harmonics. At this point, one should
observe that%
\begin{equation}
Y_{\tilde{q}\ell m}\left( \theta ,\varphi \right) =\left\{ 
\begin{tabular}{ll}
$e^{i\left( m+q\right) \varphi }\,\left( 1-x\right) ^{\sigma/2}\left( 1+x\right)
^{\nu/2}\,P_{q\ell m}\left( x\right) ;\;$ & in region $R_{A}$ \\ 
$e^{i\left( m-q\right) \varphi }\,\left( 1-x\right) ^{\sigma/2}\left( 1+x\right)
^{\nu/2}\,P_{q\ell m}\left( x\right) ;$ & in region $R_{B}$%
\end{tabular}%
\right. .  \label{e29}
\end{equation}%
\begin{figure}[!ht]  
\centering
\includegraphics[width=0.3\textwidth]{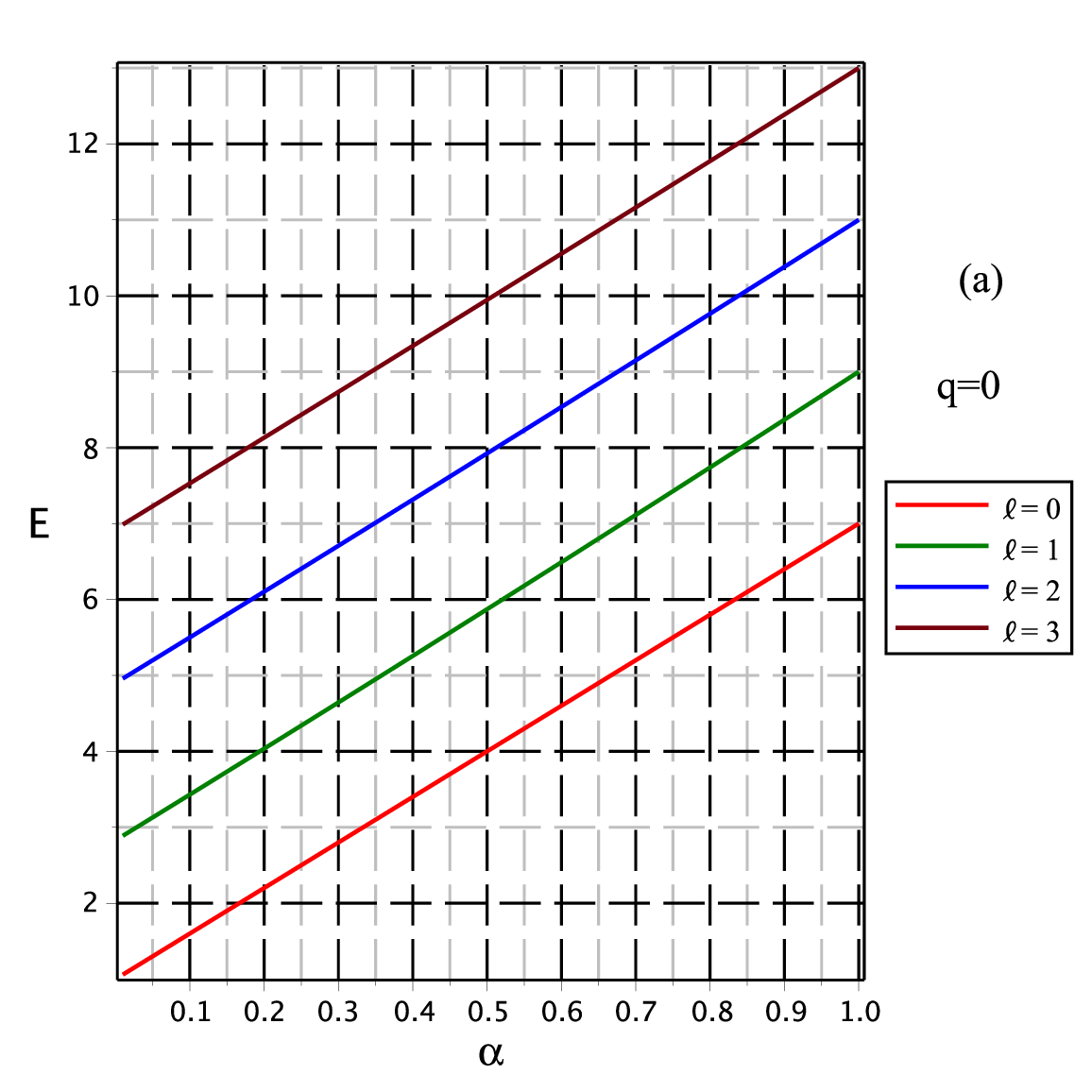}
\includegraphics[width=0.3\textwidth]{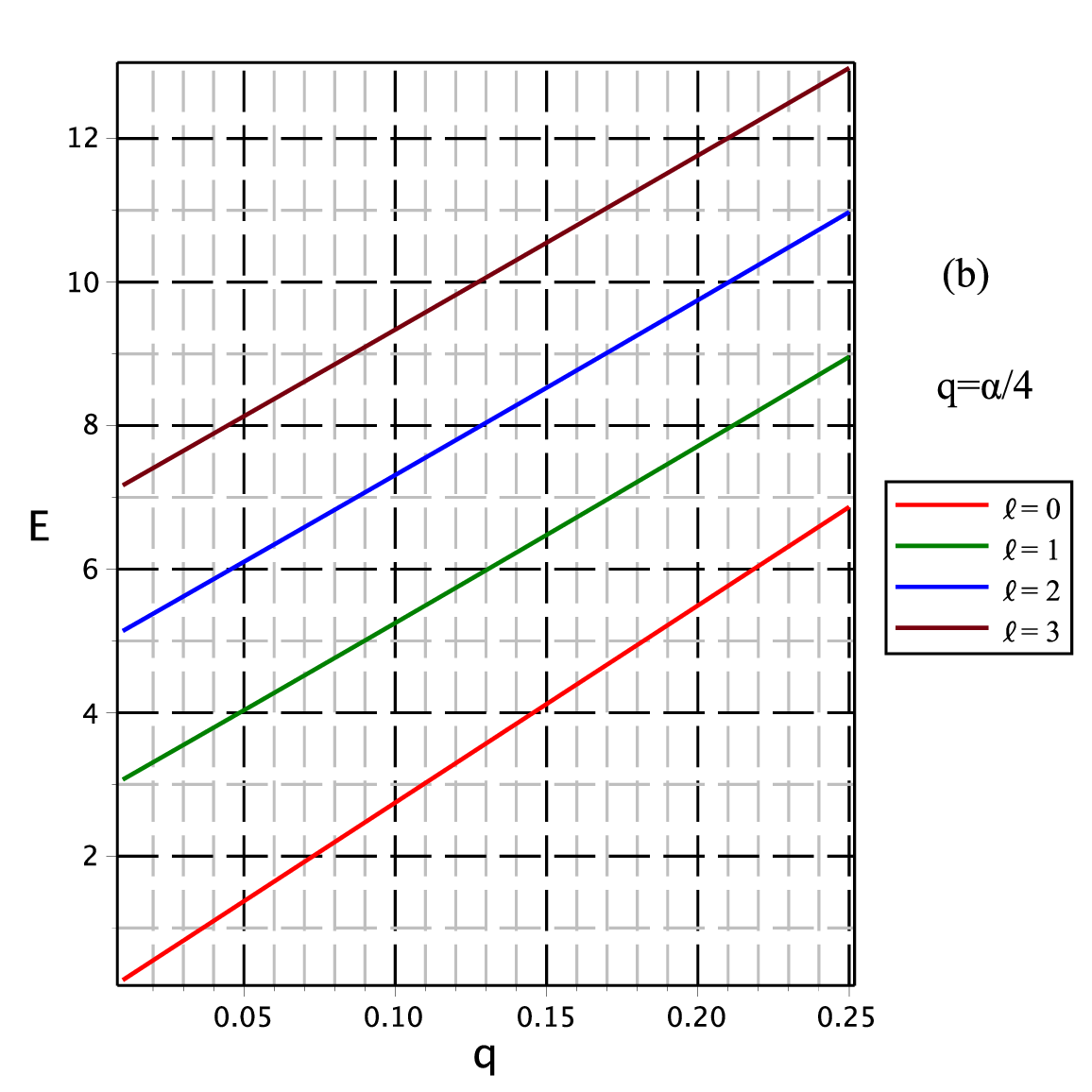} 
\includegraphics[width=0.3\textwidth]{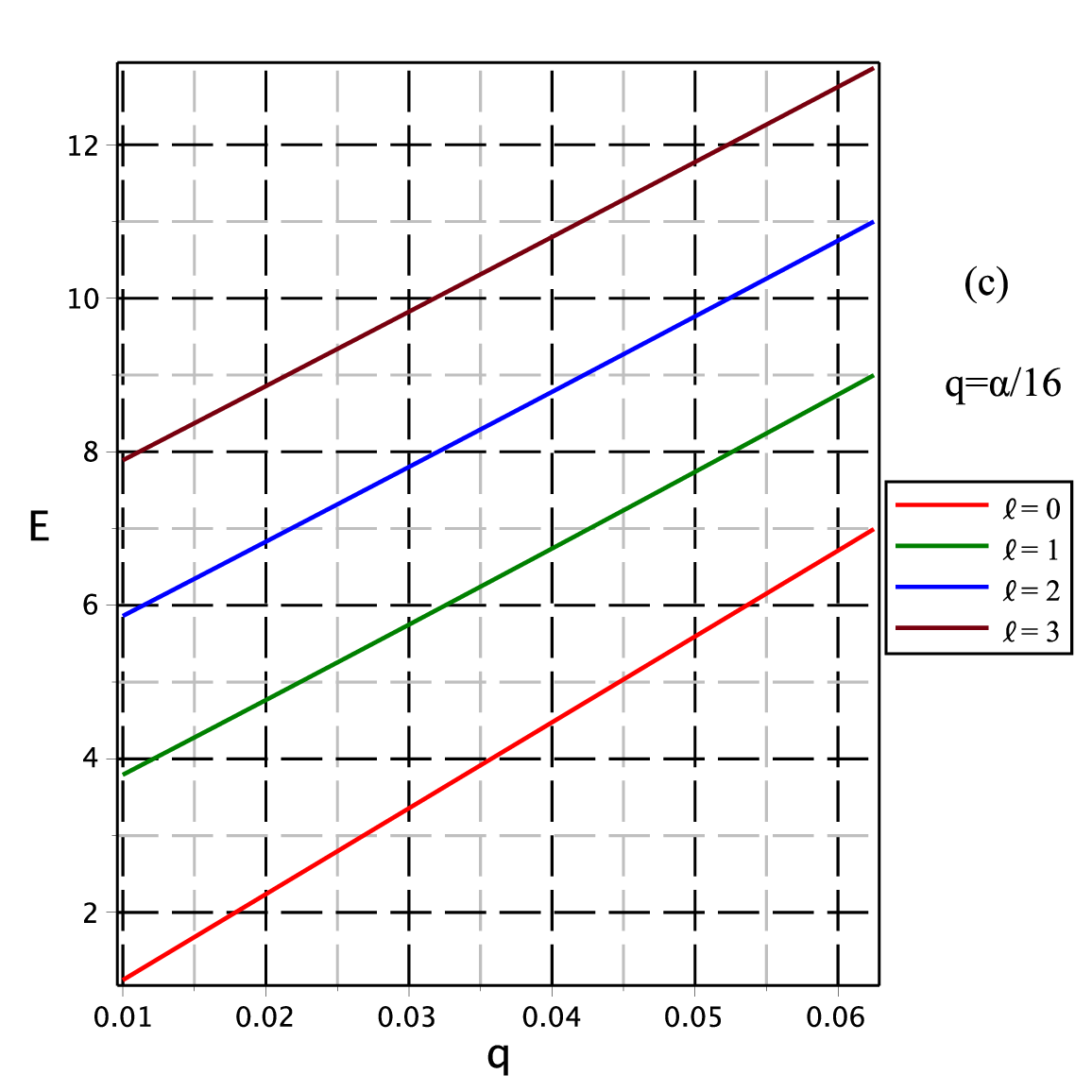}
\caption{\small 
{ The energy levels, Eq. (\ref{e32}),  of the PDM\ Schr\"{o}%
dinger oscillators in a PGM background and a Wu-Yang magnetic monopole for $%
n_{r}=1$, $\ell =0,1,2,3$, $\omega =1$, and (a) $q=0$ (i.e., no Wu-Yang
magnetic monopole), (b) $q=\alpha /4$, and (c) $q=\alpha /16$.}}
\label{fig1}
\end{figure}%
We may now rewrite the radial equation (\ref{e22}), with\ $V\left( \mathbf{%
\rho }\left( r\right) \right) =\omega ^{2}r^{2}$,\ as%
\begin{equation}
\left\{ -\frac{1}{r^{2}}\,\partial _{r}\left( r^{2}\,\partial _{r}\right) +%
\frac{L\left( L+1\right) }{r^{2}}+\tilde{\omega}^{2}r^{2}\right\} \psi
\left( \rho \left( r\right) \right) =\mathcal{E}\psi \left( \rho \left(
r\right) \right) ,  \label{e30}
\end{equation}%
where $\mathcal{E}=E/\alpha ^{2}$,$\;\tilde{\omega}=\omega /\alpha $ and%
\begin{equation}
L\left( L+1\right) =\frac{\ell \left( \ell +1\right) -q^{2}}{\alpha ^{2}}%
\Longrightarrow L=-\frac{1}{2}+\sqrt{\frac{1}{4}+\frac{\ell \left( \ell
+1\right) -q^{2}}{\alpha ^{2}}}.  \label{e31}
\end{equation}%
One should notice that the square root signature is chosen so that for $%
\alpha =1$ and $q=0$ one would retrieve $L=\ell $ the regular angular
momentum quantum number. Moreover, the solution to (\ref{e30}) with $\psi
\left( \rho \right) =R\left( \rho \right) /\rho $ would read%
\begin{equation}
R\left( r\right) =R\left( r\left( \rho \right) \right) \sim r^{L+1}\exp
\left( -\frac{\tilde{\omega}r^{2}}{2}\right) \;_{1}F_{1}\left( \frac{L}{2}+%
\frac{3}{4}-\frac{\mathcal{E}}{4\tilde{\omega}},L+\frac{3}{2},\tilde{\omega}%
r^{2}\right) ;\;r=\sqrt{q\left( \rho \right) }\rho .  \label{e311}
\end{equation}%
However, finiteness and square integrability would again enforce the
condition that the confluent hypergeometric series is truncated into a
polynomial of order $n_{r}=0,1,2,\cdots $ so that $\frac{L}{2}+\frac{3}{4}-%
\frac{\mathcal{E}}{4\tilde{\omega}}=-n_{r}$ to imply 
\begin{equation}
\mathcal{E}=2\tilde{\omega}\left( 2n_{r}+L+\frac{3}{2}\right) \Rightarrow
E_{n_{r},\ell ,q}=2\alpha \omega \left( 2n_{r}+\sqrt{\frac{1}{4}+\frac{\ell
\left( \ell +1\right) -q^{2}}{\alpha ^{2}}}+1\right) .  \label{e32}
\end{equation}%
The Schr\"{o}dinger oscillators described in (\ref{e30}) are isospectral and
invariant with the corresponding PDM Schr\"{o}dinger oscillators%
\begin{equation}
\left\{ --\frac{1}{q\left( \rho \right) \,\sqrt{f\left( \rho \right) }\,\rho
^{2}}\,\partial _{\rho }\left( \frac{q\left( \rho \right) \,\,\rho ^{2}}{%
\sqrt{f\left( \rho \right) }}\,\partial _{\rho }\right) +\frac{L\left(
L+1\right) }{q\left( \rho \right) \,\rho ^{2}}+\tilde{\omega}^{2}q\left(
\rho \right) \,\rho ^{2}\right\} \psi \left( \rho \right) =\mathcal{E}\psi
\left( \rho \right) ,  \label{e33}
\end{equation}%
\begin{figure}[!ht]  
\centering
\includegraphics[width=0.3\textwidth]{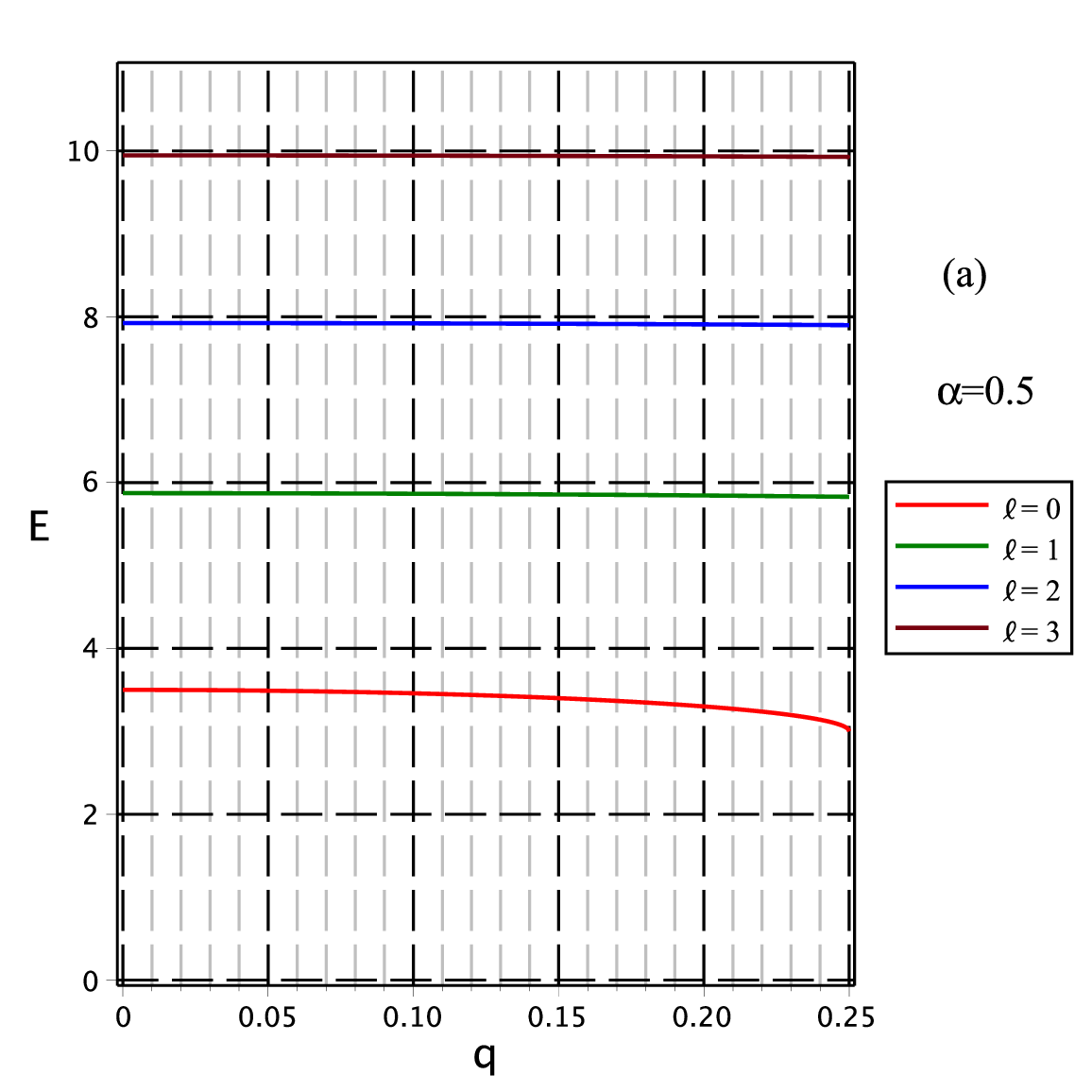}
\includegraphics[width=0.3\textwidth]{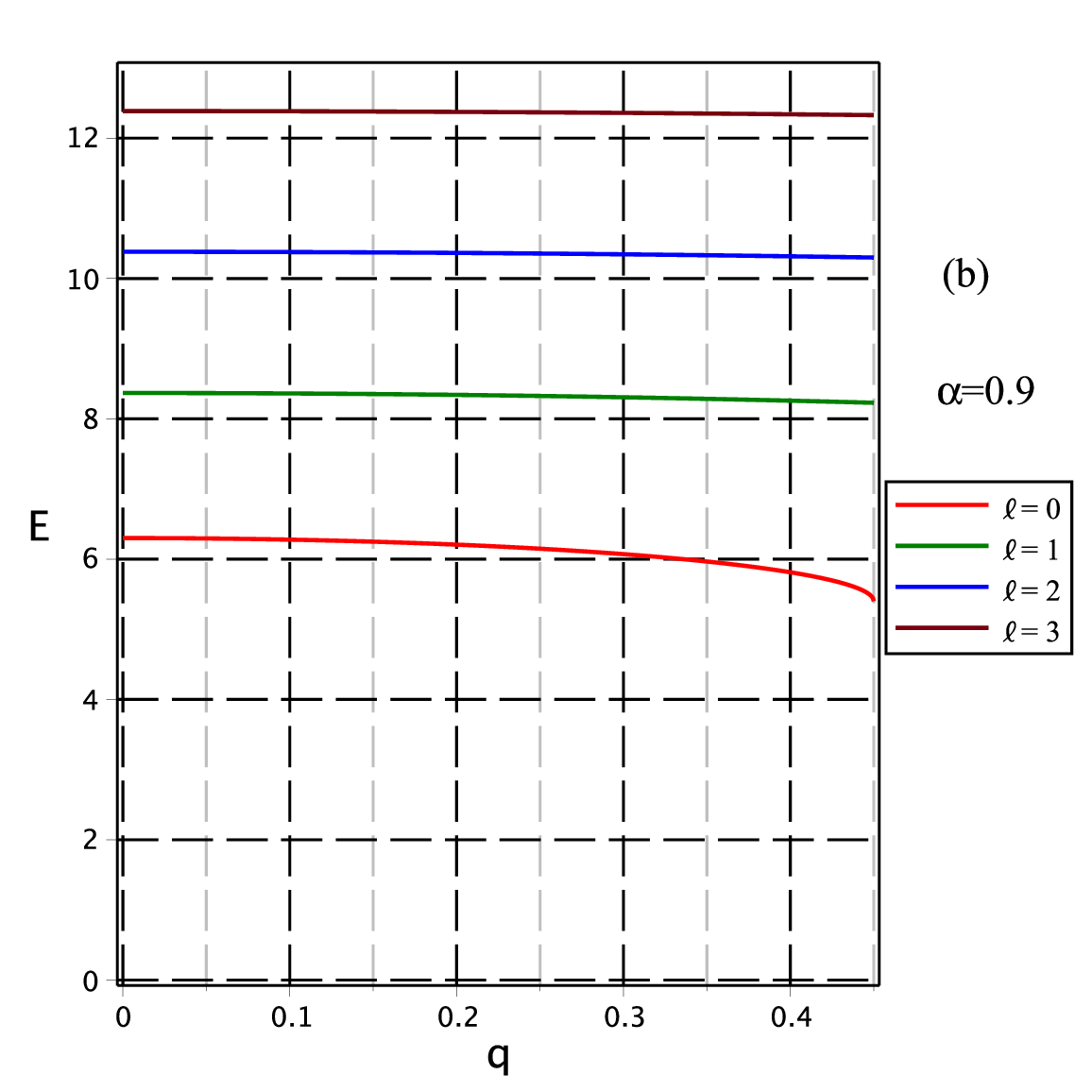} 
\includegraphics[width=0.3\textwidth]{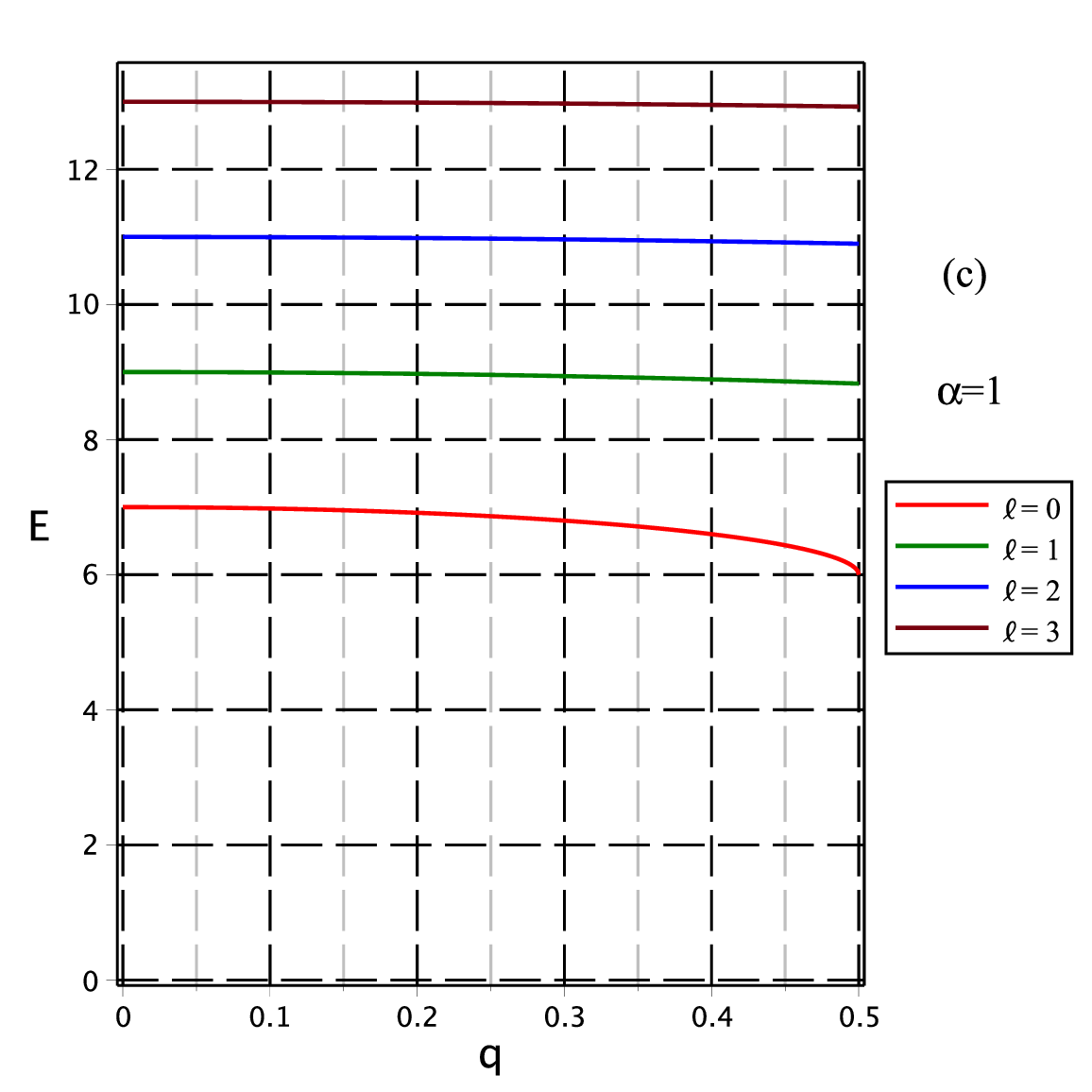}
\caption{\small 
{ The energy levels, Eq. (\ref{e32}), of the PDM\ Schr\"{o}%
dinger oscillators in a PGM background and a Wu-Yang magnetic monopole for $%
n_{r}=1$, $\ell =0,1,2,3$, $\omega =1$, and for different values of Wu-Yang
magnetic monopole parameter $q=eg$ at  (a) $\alpha =0.5$, (b) $\alpha =0.9$,
and (c) $\alpha =1$ (i.e., flat Minkowski spacetime).}}
\label{fig2}
\end{figure}%
At this point, we may report that the energy levels of (\ref{e32}) are
plotted in Figures 1 and 2.

In Figures 1(a), 1(b), and 1(c), we show the energy levels of (\ref{e32})
against the global monopole parameter $\alpha $ for different Wu-Yang
magnetic monopole parameter values $q=0$, $q=\alpha /4$, and $q=\alpha /16$,
respectively. For $q=0$ (i.e., No Wu-Yang monopole), we observe in Figure
1(a) that while the energy levels linearly increase with increasing $\alpha 
$, the spacing between the energy levels (for the same $n_{r}$ and $\ell
=0,1,2,3$, where $n_{r}=1$ is used throughout) remains constant at each $%
\alpha $ value. This is a common characteristic for the Schr\"{o}dinger
oscillator in a flat Minkowski spacetime (i.e., $\alpha =1$). However, in
1(b) and 1(c) ( for $q=\alpha /4$, and $q=\alpha /16$, respectively), we
notice that the equal spacing between energy levels is no longer valid. The
maximum value for $q$ used are chosen so that $\alpha _{\max }=1$. In
Figures 2(a), 2(b), and 2(c), we show \ the energy levels at $\alpha =0.5$, $%
\alpha =0.9$, and $\alpha =1$, respectively, for different Wu-Yang magnetic
monopole strengths $q=eg$. Where the maximum values for $q$ are now chosen
so that the square root in (\ref{e32}) remains a real-valued one. We observe
that the Wu-Yang monopole yields non-equally spaced energy levels. Moreover, it is clear that the energies are shifted up as the PGM parameter $\alpha$ increases for each value of the Wu-Yang monopole parameter $q$ (including $q=0$ for no Wu-Yang monopole).

\section{Thermodynamical properties of the PDM\ Schr\"{o}dinger oscillators
in a PGM background and a Wu-Yang magnetic monopole}

In this section we shall study the thermodynamical properties of PDM\ Schr%
\"{o}dinger oscillators in a global monopole spacetime background without
and with a Wu-Yang magnetic monopole. In a straightforward manner one
obtains the partition function%
\begin{equation}
Z\left( \beta \right) =\sum\limits_{n_{r}=0}^{\infty }\exp \left( -\beta
\,E_{n_{r},\ell ,q}\right) =\frac{\exp \left( -2\alpha \beta \omega \tau
\right) }{1-\exp \left( -4\alpha \beta \omega \right) };\text{ }\beta =\frac{%
1}{K_{B}T},  \label{e34}
\end{equation}%
where $K_{B}$ is the Boltzmann constant, $T$ is the temperature and 
\begin{equation}
\tau =1+\frac{1}{2\alpha }\sqrt{\alpha ^{2}+4\ell \left( \ell +1\right)
-4q^{2}}.  \label{e341}
\end{equation}%
\begin{figure}[!ht]  
\centering
\includegraphics[width=0.3\textwidth]{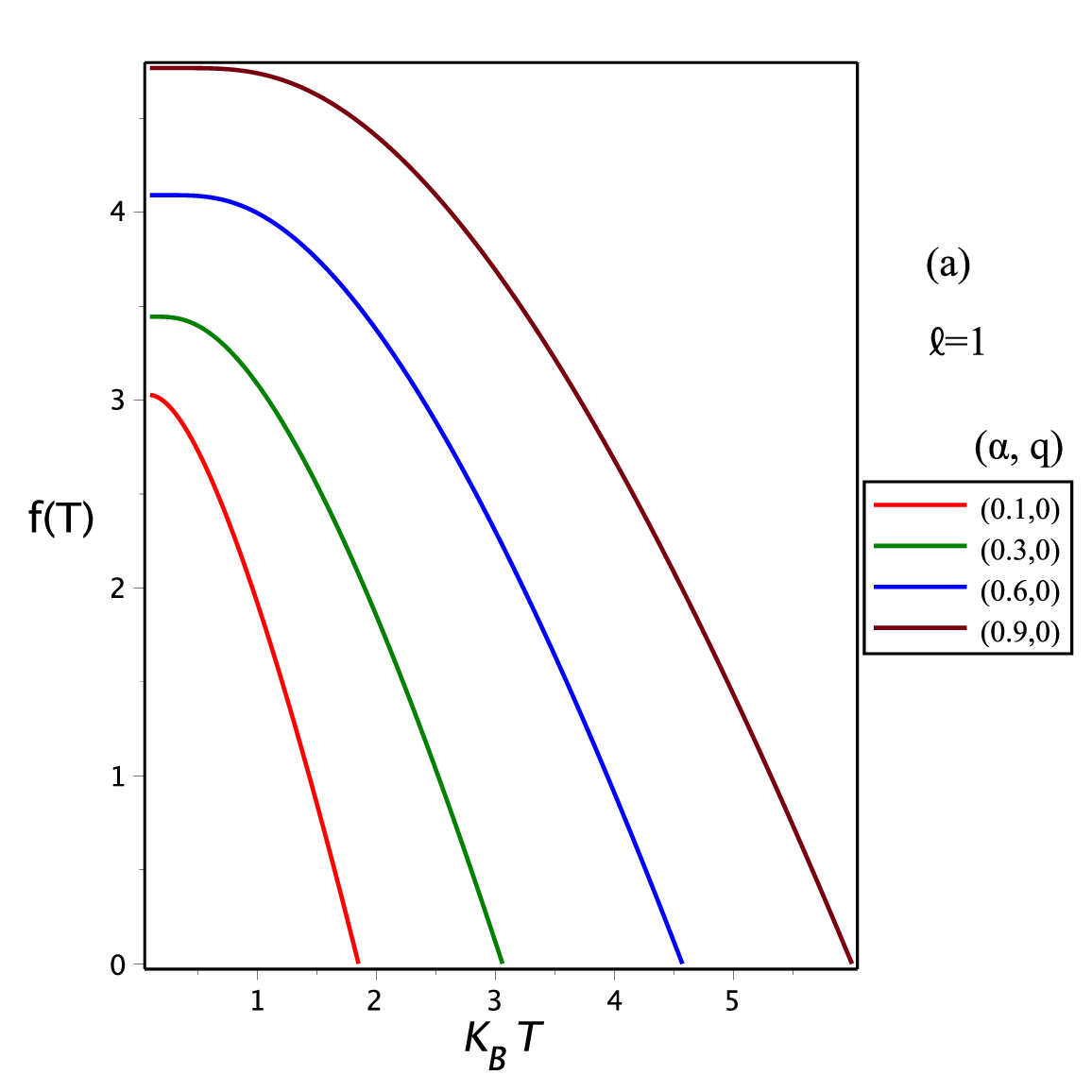}
\includegraphics[width=0.3\textwidth]{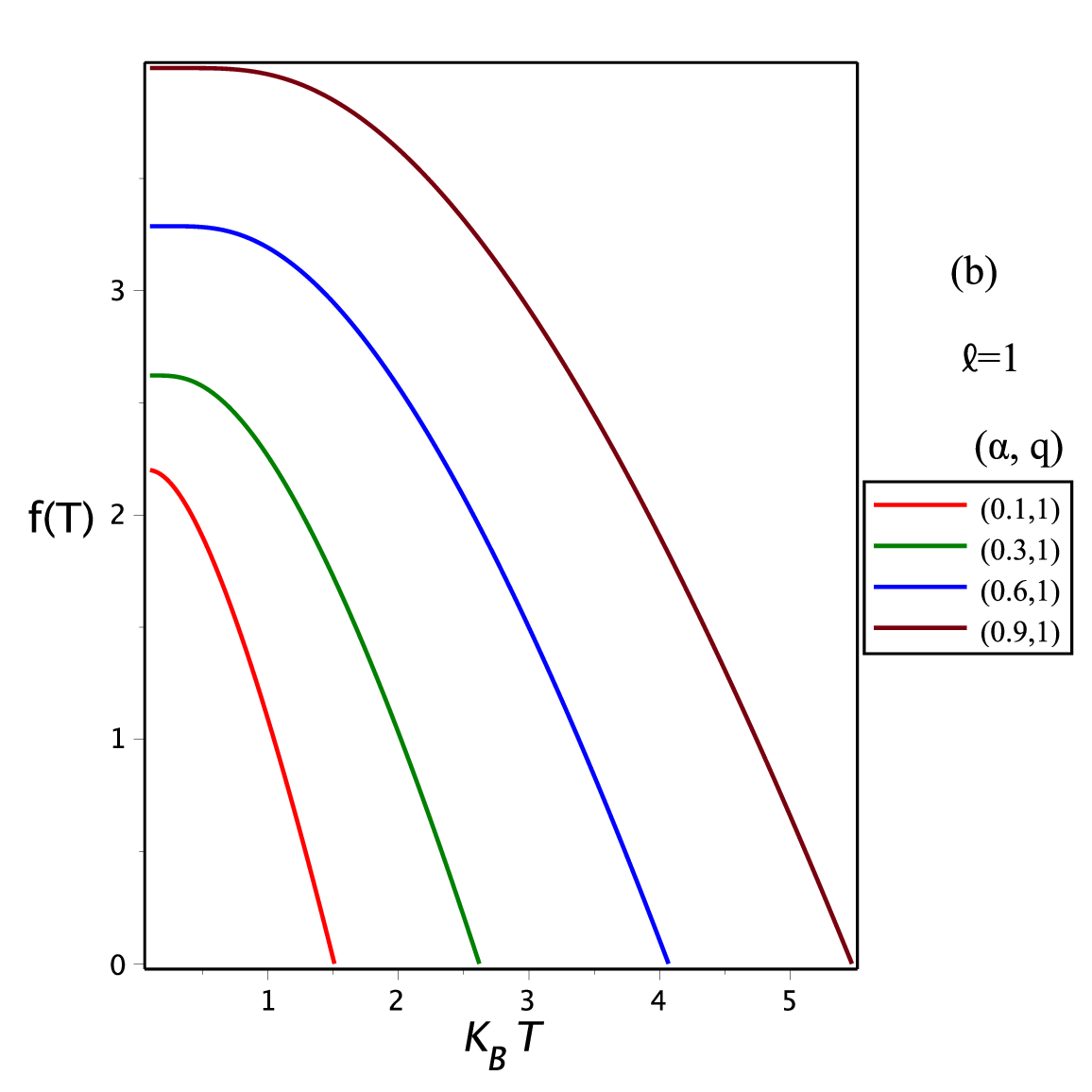} 
\includegraphics[width=0.3\textwidth]{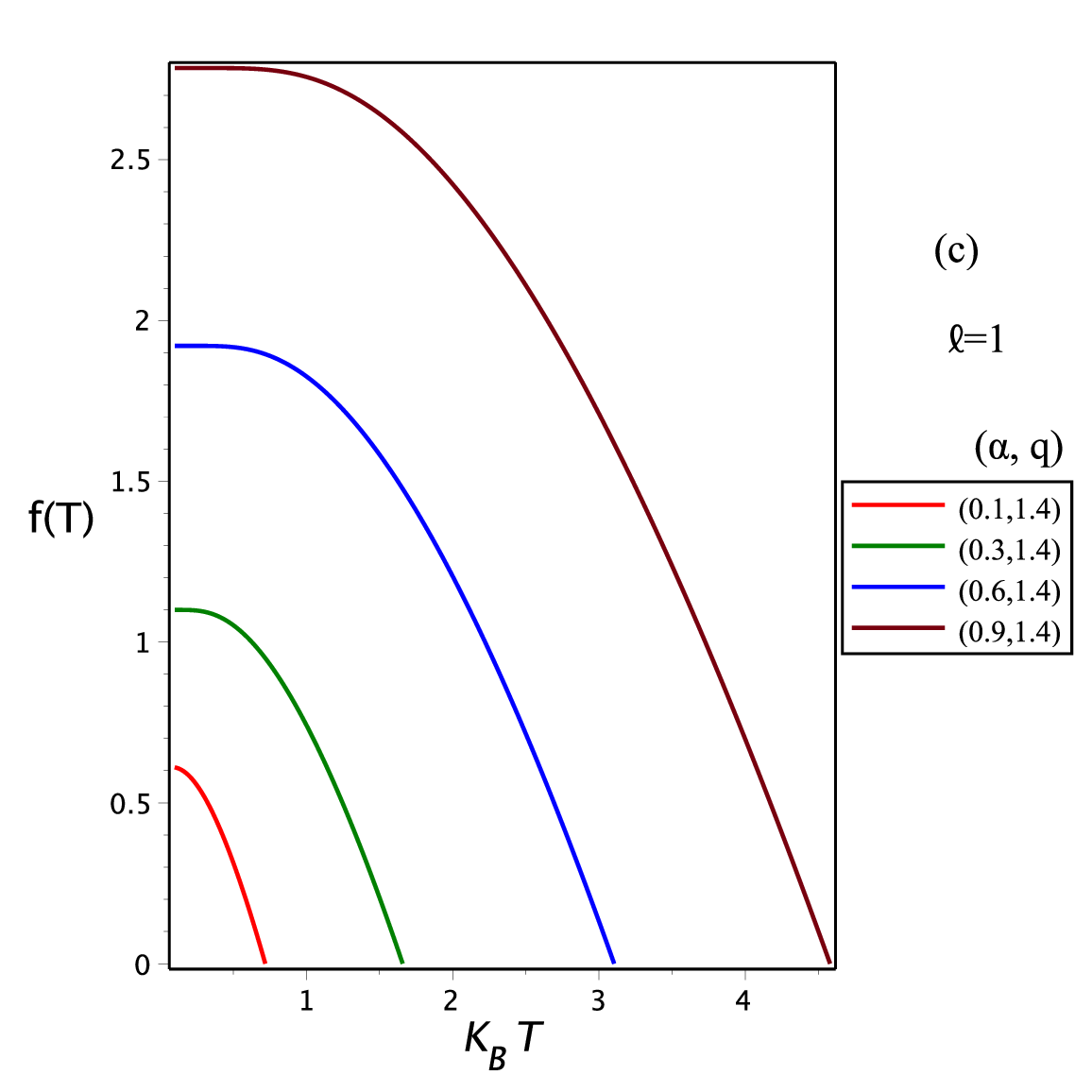}
\caption{\small 
{ The Helmholtz free energies $f\left( T\right) $, (\ref{e35}),
against $K_{B}T$ of the PDM\ Schr\"{o}dinger oscillators in a PGM background
and a Wu-Yang magnetic monopole for $\ell =1$, $\omega =1$, and $\alpha
=0.1,0.3,0.6,0.9$\ at (a) $q=0$, (b) $q=1$, and (c) $q=1.4$.}}
\label{fig3}
\end{figure}%
At this point, one should notice that for $q=eg=0$ represents PDM Schr\"{o}%
dinger oscillators in a global monopole spacetime background without the
Wu-Yang magnetic monopole. Moreover, the global monopole parameter $\alpha $
and the Wu-Yang monopole strength (through $q=eg$) are correlated in such a
way that the value under the square root remains real. In this case, $0\leq
q\leq \sqrt{\ell \left( \ell +1\right) +\alpha ^{2}/4}$,  and consequently $q _{\max }=\ell +1/2$, where $\alpha _{\max }=1$ corresponds to flat Minkowski spacetime.

To observe the effects of the global monopole spacetime background and the
Wu-Yang magnetic monopole on some thermodynamical properties associated with
such systems, we find that the Helmholtz free energy $f\left( T\right) $ is
given by
\begin{equation}
f\left( T\right) =-\frac{1}{\beta }\ln \left( Z\left( \beta \right) \right)
=2\alpha \omega \tau +K_{B}T\,\ln \left( 1-\exp \left( -\frac{4\alpha \omega 
}{K_{B}T}\right) \right) ,  \label{e35}
\end{equation}%
the Entropy $S\left( T\right) $
\begin{equation}
S\left( T\right) =-\frac{df\left( T\right) }{dT}=-K_{B}\,\ln \left( 1-\exp
\left( -\frac{4\alpha \omega }{K_{B}T}\right) \right) +\frac{4\alpha \omega 
}{T}\left[ \frac{\exp \left( -\frac{4\alpha \omega }{K_{B}T}\right) }{1-\exp
\left( -\frac{4\alpha \omega }{K_{B}T}\right) }\right] ,  \label{e36}
\end{equation}%
the Specific heat $c\left( T\right) $

\begin{equation}
c\left( T\right) =T\,\frac{dS\left( T\right) }{dT}=\frac{16\alpha ^{2}\omega
^{2}}{T^{2}K_{B}}\left[ \frac{\exp \left( -\frac{2\alpha \omega }{K_{B}T}%
\right) }{1-\exp \left( -\frac{4\alpha \omega }{K_{B}T}\right) }\right] ^{2},
\label{e37}
\end{equation}%
and Mean energy $U\left( T\right) $

\begin{equation}
U\left( T\right) =-\frac{dZ\left( \beta \right) }{d\beta }=2\alpha \omega
\tau -\frac{4\alpha \omega }{1-\exp \left( \frac{4\alpha \omega }{K_{B}T}%
\right) }.  \label{e38}
\end{equation}%
We observe that while the Helmholtz free energy $f\left( T\right) $ in (\ref%
{e35}) and the Mean energy $U\left( T\right) $ in (\ref{e38}) are affected by the Wu-Yang magnetic monopole through the parameter $\tau $ in (\ref{e341}%
), the Entropy $S\left( T\right) $ in (\ref{e36}) and the Specific heat $%
c\left( T\right) $\ in (\ref{e37}) are not. However, all mentioned
thermodynamical properties are affected by the global monopole through the
parameter $\alpha $.
\begin{figure}[!ht]  
\centering
\includegraphics[width=0.3\textwidth]{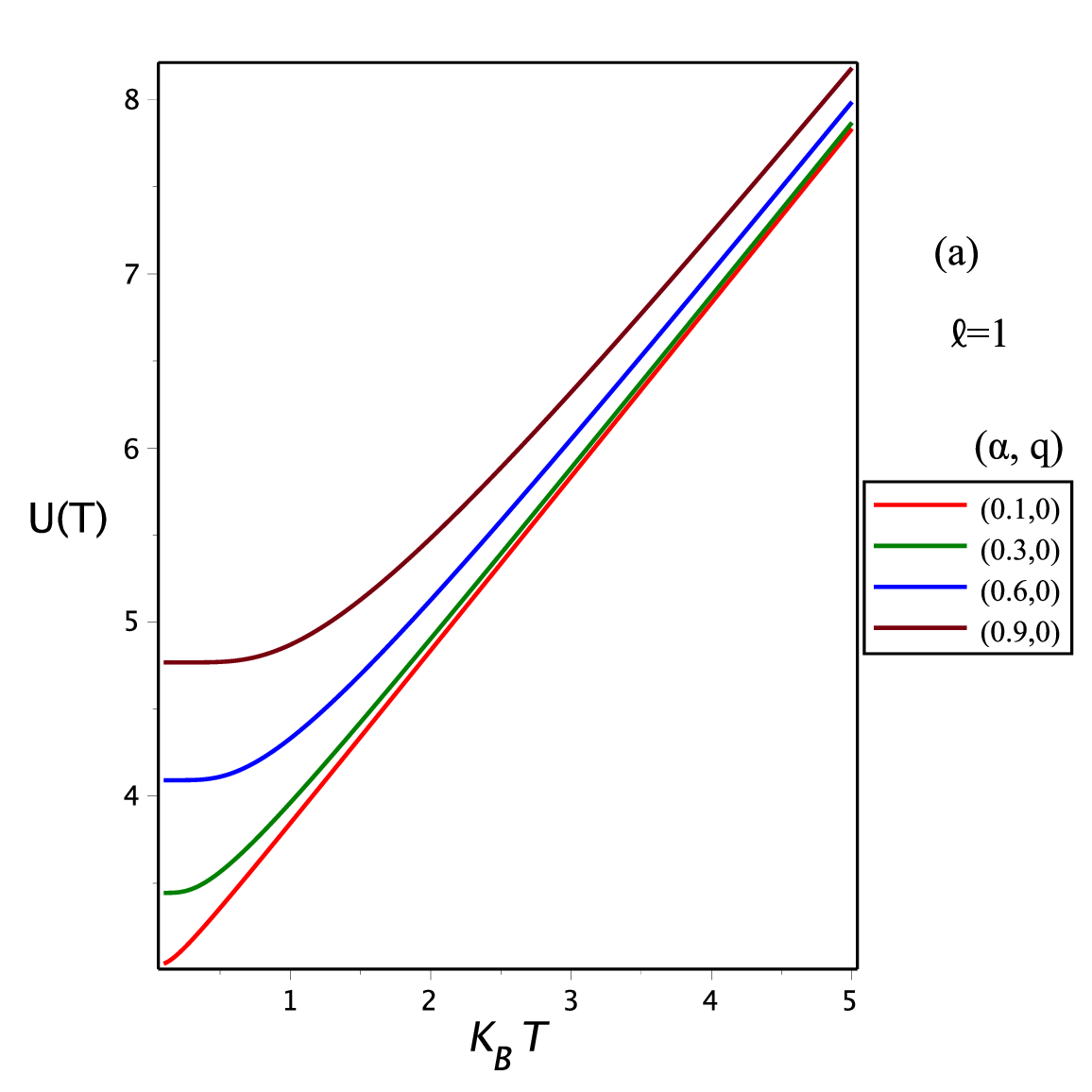}
\includegraphics[width=0.3\textwidth]{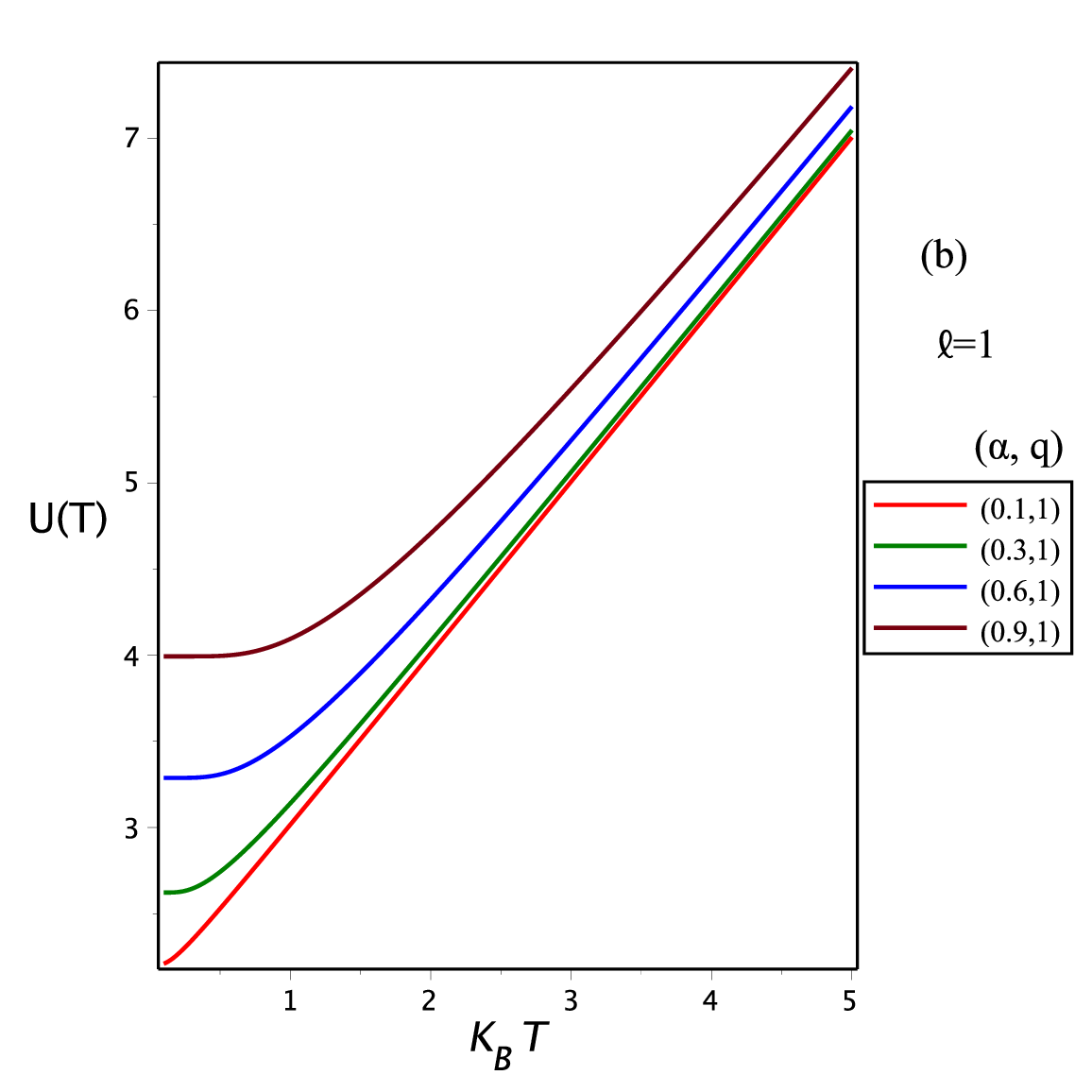} 
\includegraphics[width=0.3\textwidth]{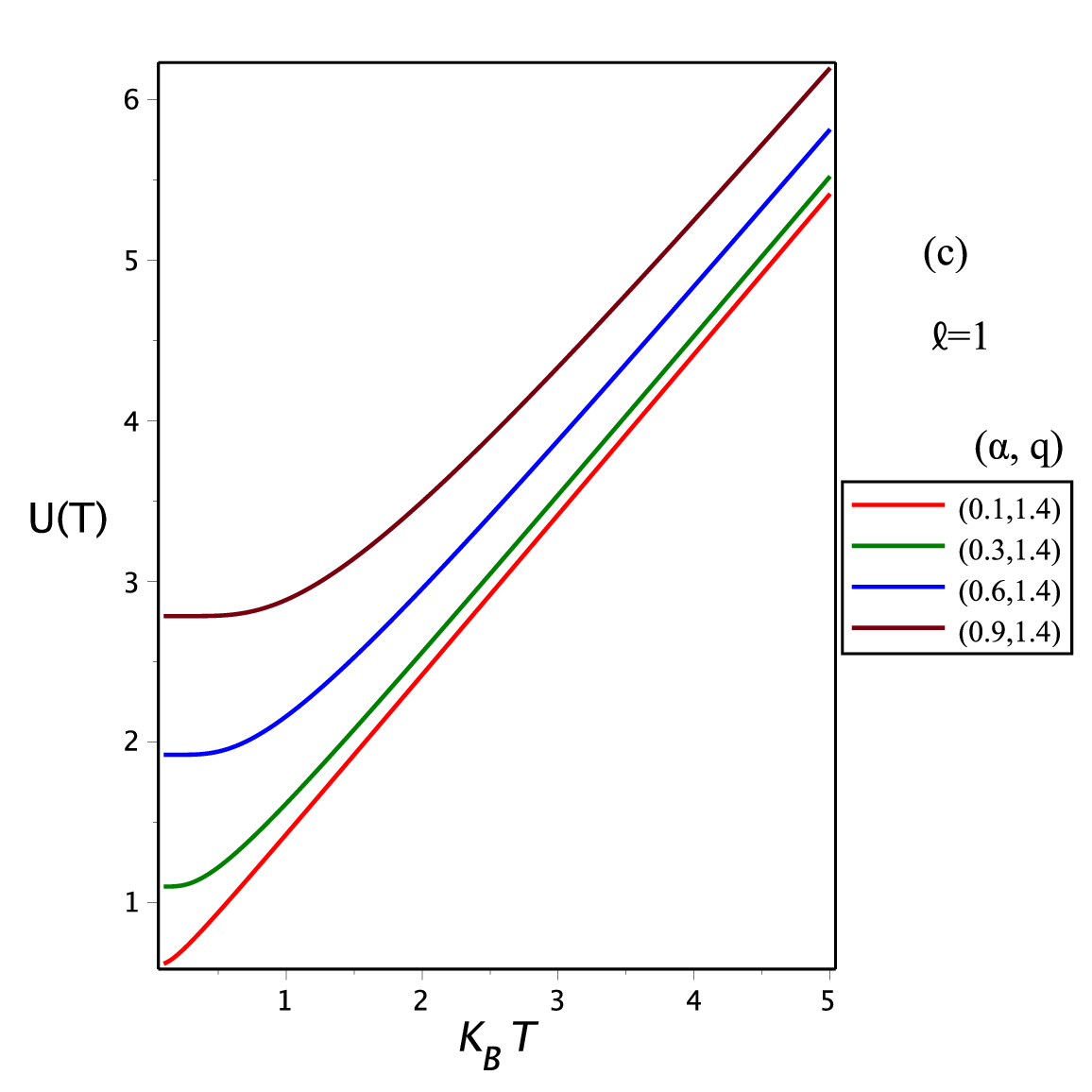}
\caption{\small 
{ The mean energies $U\left( T\right) $, Eq. (\ref{e38}),
against  $K_{B}T$ of the PDM\ Schr\"{o}dinger oscillators in a PGM
background and a Wu-Yang magnetic monopole for $\ell =1$, $\omega =1$, and $%
\alpha =0.1,0.3,0.6,0.9$\ at (a) $q=0$, (b) $q=1$,  and (c) $q=1.4$.}}
\label{fig4}
\end{figure}%
\begin{figure}[!ht]  
\centering
\includegraphics[width=0.3\textwidth]{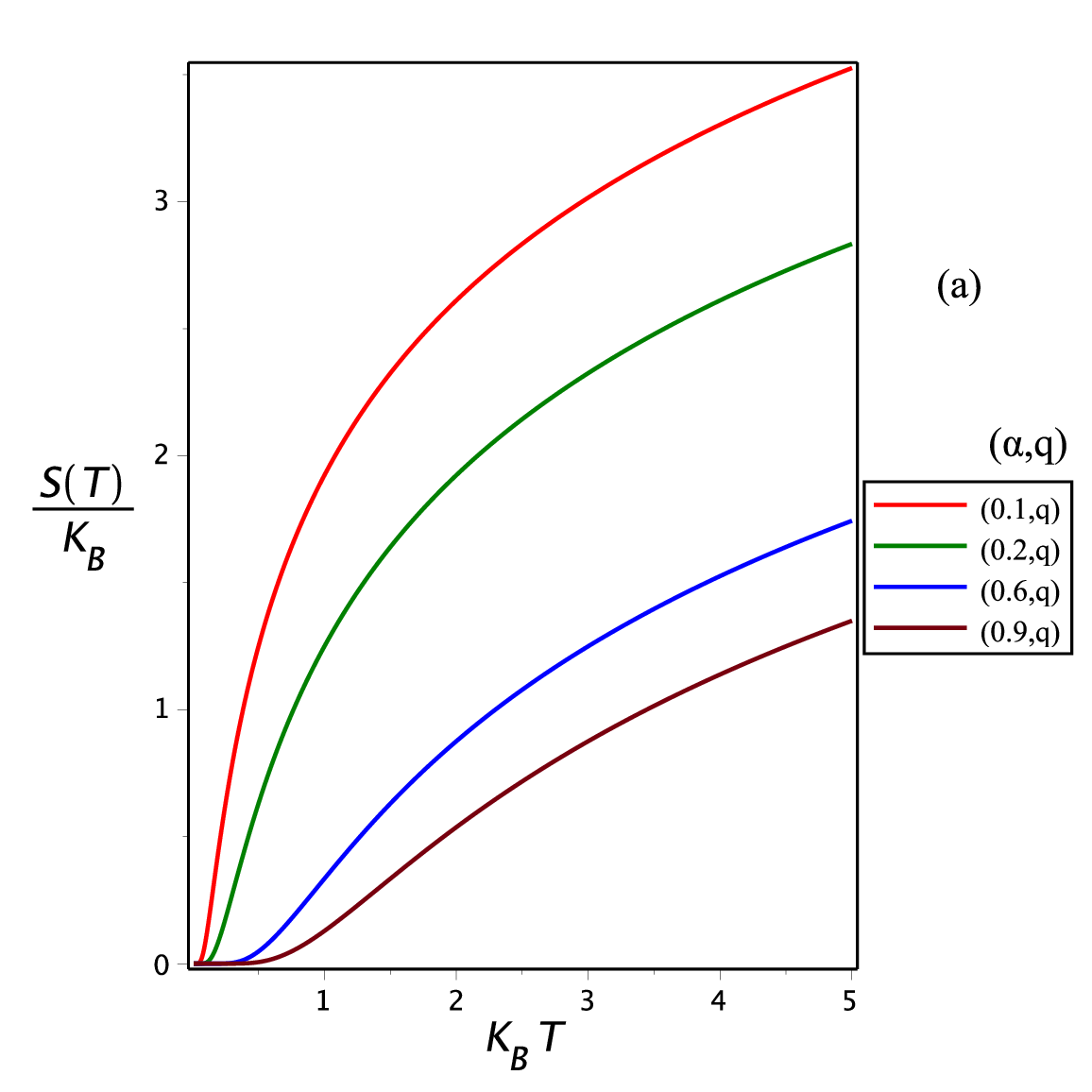}
\includegraphics[width=0.3\textwidth]{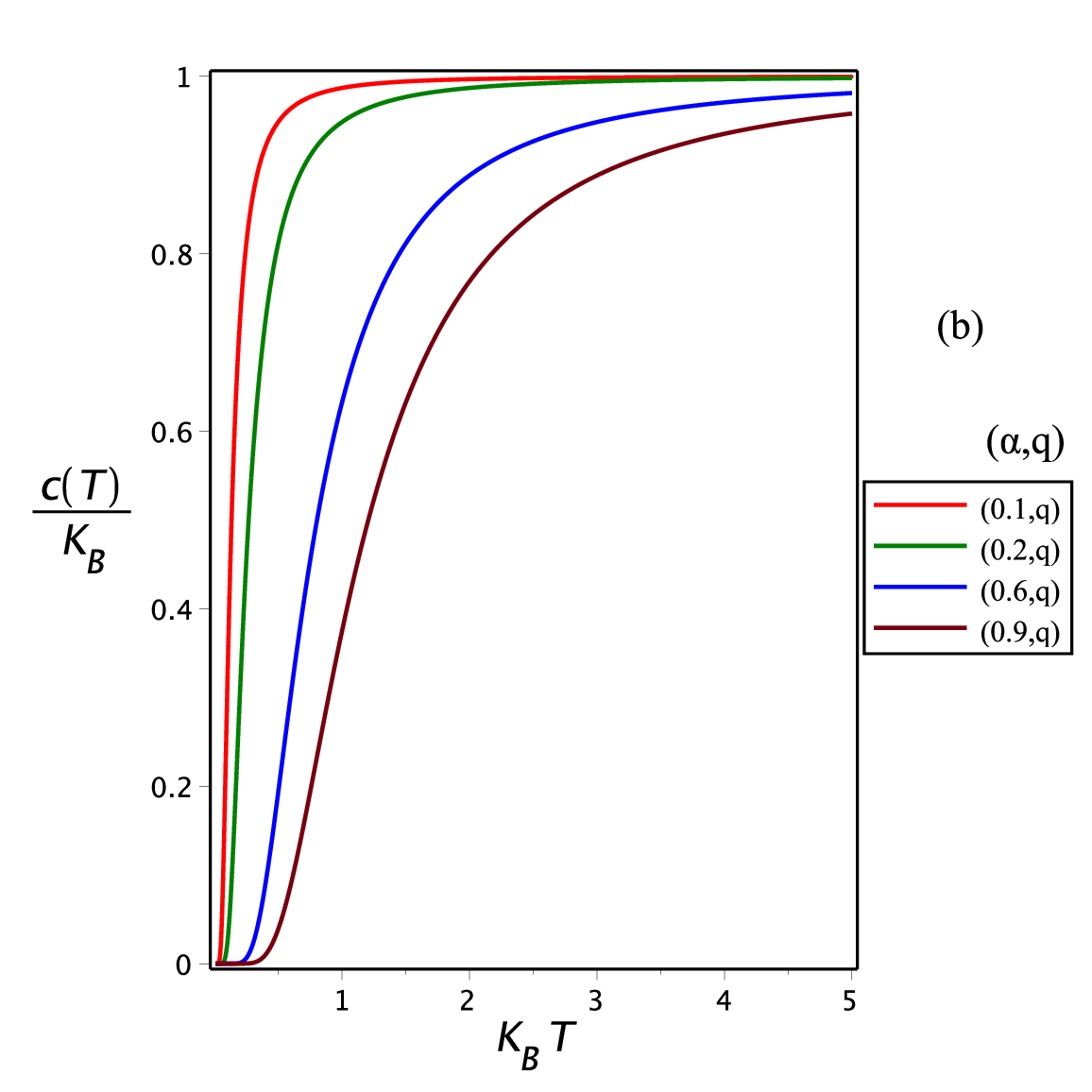} 
\caption{\small 
{ For $\ell =1$, $\omega =1$, $\alpha =0.1,0.2,0.6,0.9$\ at all
values of $q$ (i.e., the Wu-Yang magnetic monopole has no effect on the
Entropy) we show (a) the ratio $S\left( T\right) /K_{B}$, where $S\left(
T\right) $ is the Entropy, against $K_{B}T$ of the PDM\ Schr\"{o}dinger
oscillators in a PGM background and a Wu-Yang magnetic monopole, and (b) The
ratio $c\left( T\right) /K_{B}$, where $c\left( T\right) $ is the Specific
heat, against $K_{B}T$ of the PDM\ Schr\"{o}dinger oscillators in a PGM
background and a Wu-Yang magnetic monopole.}}
\label{fig5}
\end{figure}%

In Figures 3(a), 3(b), and 3(c), we show (for $\ell =1$ states) the effect
of the Wu-Yang \cite{Re101} magnetic monopole on the Helmholtz free energies 
$f(T)$, Eq.(\ref{e35}), of the Schr\"{o}dinger-oscillator in a point-like
global monopole for $q=0$, $q=1$, and $q=1.4$, respectively. It is obvious
that as $q=eg$ increases the Helmholtz free energy converges more rapidly to the zero
value as the temperature $T$ grows up from just above zero. In Figures 4(a),
4(b), and 4(c), we show (for $\ell =1$ states) the effect of the Wu-Yang
monopole on the mean energy $U\left( T\right) $, Eq. (\ref{e38}), for $q=0$, 
$q=1$, and $q=1.4$, respectively. We observe that as the Wu-Yang monopole
strength increases (through $q=eg$) the mean energy decreases for each value
of $T$. We also notice that the mean energy $U\left( T\right) $, for all
allowed $\alpha $ values used, tend to cluster at very high temperatures for $%
q=0$ (i.e., no Wu-yang monopole). However, it is clear that as $q$ increases
from zero, such clustering is slowed down. In Figure 5(a), we show (for $%
\ell =1$ states) the entropy $S\left( T\right) $, Eq. (\ref{e36}), as the
temperature grows up from just above zero for the Schr\"{o}dinger-oscillator
in a point-like global monopole. Figure 5(b) shows the specific heat $%
c\left( T\right) $, Eq. (\ref{e37}), against the temperature for the Schr%
\"{o}dinger-oscillator in a point-like global monopole. It is obvious that
the ratio $c\left( T\right) /K_{B}\rightarrow 1\Rightarrow c\left( T\right)
\rightarrow K_{B}$ as $T>>1$ for all allowed values of the point-like global
monopole parameter $\alpha $. Notably, the Wu-Yang magnetic monopole has no effect on
the entropy $S\left( T\right) $ or the specific heat $c\left( T\right) $ as
the results in (\ref{e36}) and (\ref{e37}), respectively, suggest. 
\begin{figure}[!ht]  
\centering
\includegraphics[width=0.3\textwidth]{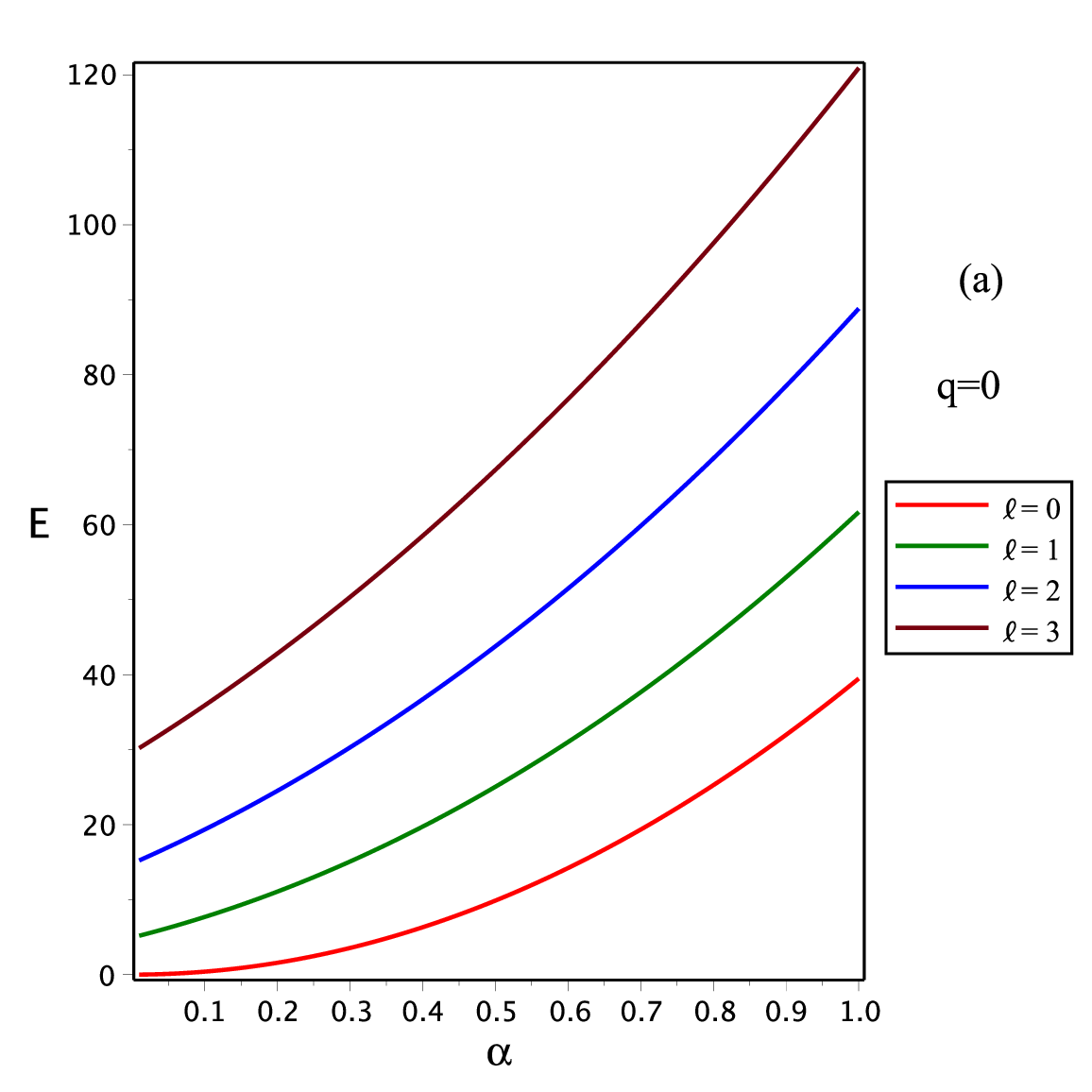}
\includegraphics[width=0.3\textwidth]{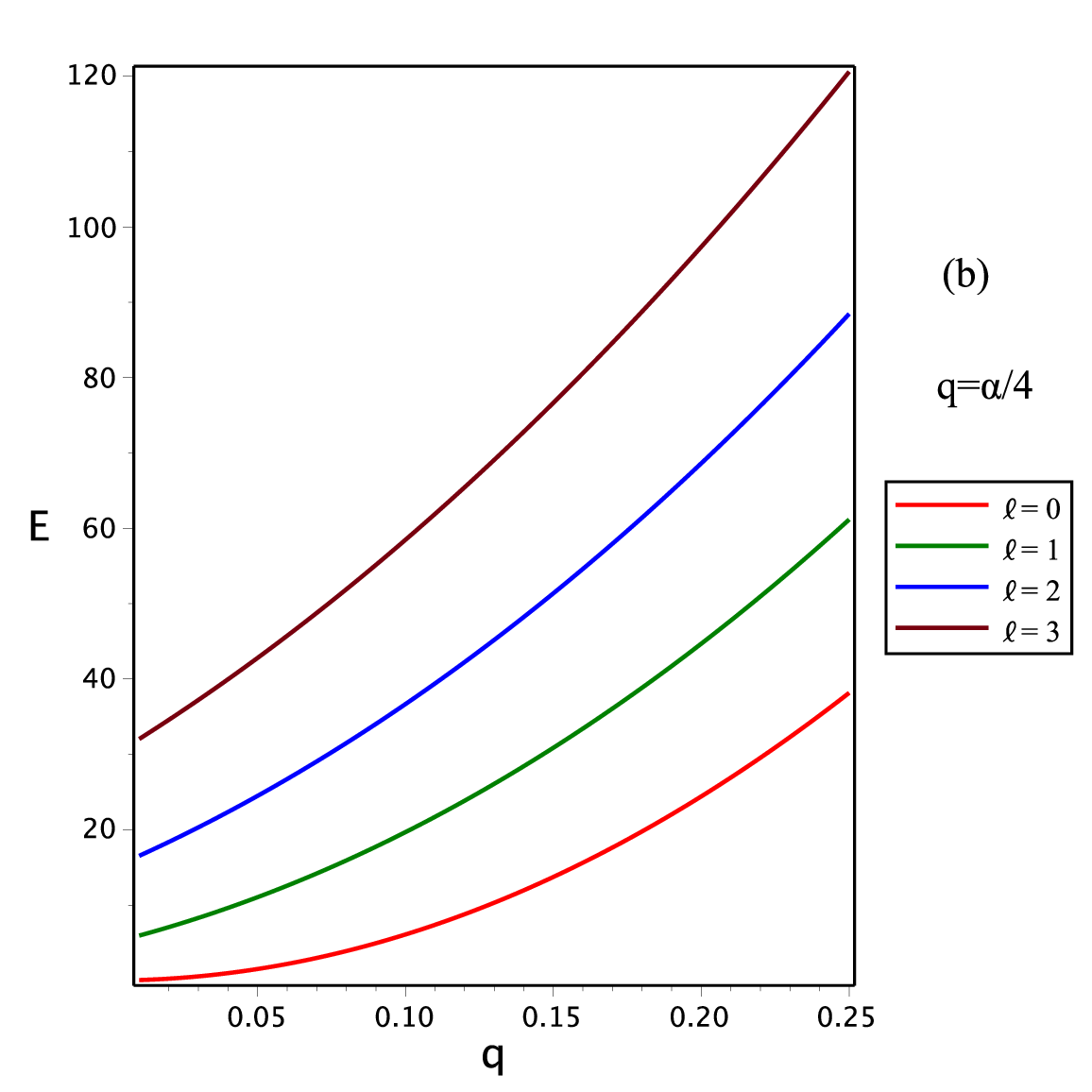} 
\includegraphics[width=0.3\textwidth]{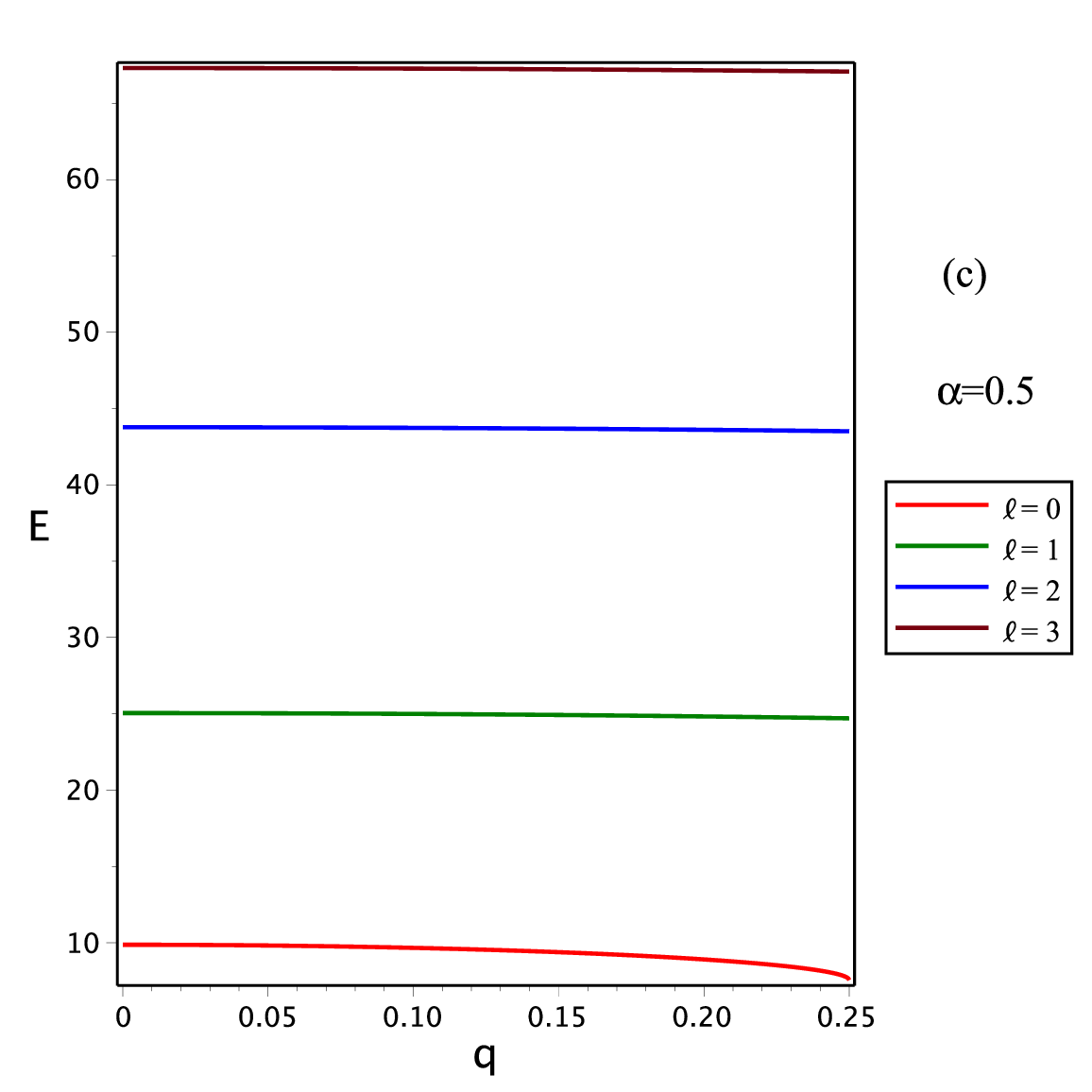}
\caption{\small 
{ We show the hard-wall, at $r=r_{\circ }=1$, effects on the
energy levels, Eq. (\ref{e41}) for $n_{r}=1$, and $\ell =0,1,2,3$. In (a)
and (b) the energy levels are plotted against the PGM parameter $\alpha $ at $%
q=0$ (i.e., no Wu-yang monopole) and $q=\alpha /4$, respectively. In (c) the
energy levels are plotted against the Wu-Yang monopole parameter $q$ at $%
\alpha =0.5$.}}
\label{fig6}
\end{figure}%
The same thermodynamical properties hold true for the PDM\ Schr\"{o}%
dinger-oscillators in a PGM spacetime and a Wu-Yang
magnetic monopole.

\section{PDM\ Schr\"{o}dinger oscillators in a PGM background and a Wu-Yang
magnetic monopole subjected to a hard-wall potential}

In this section, we consider that the system of PDM\ Schr\"{o}dinger
oscillators in a PGM background and a Wu-Yang magnetic monopole is now
subjected to an impenetrable hard-wall potential at some radial distance $%
r_{\circ }=\sqrt{q\left( \rho _{\circ }\right) }\rho _{\circ }$. This would
in turn restrict the motion of the PDM\ Schr\"{o}dinger oscillators
mentioned above to be confined within a spherical-box of radius $r_{\circ }$
with an impenetrable hard-wall. This would suggest that the the confluent
hypergeometric polynomials $\;_{1}F_{1}\left( \frac{L}{2}+\frac{3}{4}-\frac{%
\mathcal{E}}{4\tilde{\omega}},L+\frac{3}{2},\tilde{\omega}r^{2}\right) $ in (%
\ref{e311}) vanish at $r=r_{\circ }$ to consequently yield that $R\left(
r_{\circ }\right) =0$. One would then appeal to subsection 13.5 on the
asymptotic expansions and limiting forms of Abramowitz and Stegum \cite{Re36}
and recollect formula (13.5.14)%
\begin{equation}
\lim\limits_{a\rightarrow -\infty }\,_{1}F_{1}\left( a,b,x\right) =\Gamma
\left( b\right) \,e^{x/2}\,\pi ^{-1/2}\left( \frac{bx}{2}-ax\right)
^{1/4-b/2}\,\cos \left( \sqrt{\left( 2b-4a\right) x}-\frac{b}{2}\pi +\frac{%
\pi }{4}\right) \left[ 1+O\left( |\frac{b}{2}-a|^{-1/2}\right) \right] ,
\label{e39}
\end{equation}%
for a real $x$ and a bounded $b$. This formula immediately suggests that $a=%
\frac{L}{2}+\frac{3}{4}-\frac{\mathcal{E}}{4\tilde{\omega}}$, $b=L+\frac{3}{2%
}$, and $x=\tilde{\omega}r^{2}\Rightarrow x_{\circ }=\tilde{\omega}r_{\circ
}^{2}$. Consequently, only at very high energies of PDM\ Schr\"{o}dinger
oscillators and/or very small values of the PGM parameter $\alpha $ (i.e., $%
\mathcal{E}=E/\alpha ^{2}\mathcal{\longrightarrow \infty }$) one would have
a vanishing radial function at some $r=r_{\circ }$, i.e., $R\left( r_{\circ
}\right) =0$. Under such conditions, 
\begin{equation}
\cos \left( \sqrt{\left( 2b-4a\right) x_{\circ }}-\frac{b}{2}\pi +\frac{\pi 
}{4}\right) =0\Rightarrow \sqrt{\left( 2b-4a\right) x}-\frac{b}{2}\pi +\frac{%
\pi }{4}=\left( n_{r}+\frac{1}{2}\right) \pi   \label{e40}
\end{equation}%
one would, in a straightforward manner, obtain%
\begin{equation}
E_{n_{r},\ell ,q}=\frac{\pi ^{2}\alpha ^{2}}{4r_{\circ }^{2}}\left[ 2n_{r}+%
\sqrt{\frac{1}{4}+\frac{\ell \left( \ell +1\right) -q^{2}}{\alpha ^{2}}}+%
\frac{3}{2}\right] ^{2}  \label{e41}
\end{equation}%
Comparing this result with that of (\ref{e32}) we observe that the hard-wall
spherical box has indeed changed the corresponding energies for the PDM\ Schr\"{o}dinger oscillators in a PGM background and a Wu-Yang magnetic monopole.
In Figures 6(a),6(b), and 6(c), we show the hard-wall, at $r=r_{\circ }=1$,
effects on the energy levels, Eq. (\ref{e41}) for $n_{r}=1$, and $\ell
=0,1,2,3$.. In 6(a) and 6(b) the energy levels are plotted against the PGM
parameter $\alpha $ at $q=0$ (i.e., no Wu-yang monopole) and $q=\alpha /4$,
respectively. In 6(c) the energy levels are plotted against the Wu-Yang
monopole parameter $q$ at $\alpha =0.5$.

To figure out the hard-wall effects of the PDM\ Schr\"{o}dinger oscillators
in a PGM background and a Wu-Yang magnetic monopole, we compare between
Figures 1(a) and 6(a). We observe that the equidistance between the energy
levels in 1(a) is no longer valid in 6(a). The separation between the energy
levels in 6(a) quadratically increases with increasing PGM parameter $\alpha 
$ as it increases from just above the zero value. Notably, drastic shift-ups
in the energy levels are obvious as $\alpha $ grows up. The same trend of the hard-wall effect is also observed through the comparison between Figures
1(b) and 6(b). This is expected from $\alpha ^{2}$ dependence of $%
E_{n_{r},\ell ,q}$ in (\ref{e41}). However, the comparison between Figures
2(a) and 6(c), at a fixed $\alpha =0.5$, again suggests drastic shift-ups in
the energy levels, but, in this case, each energy level very slowly
decreases to a minimum value of
\begin{equation}
E_{\min }=\frac{\pi ^{2}\alpha ^{2}}{4r_{\circ }^{2}}\left(
2n_{r}+3/2\right) ^{2}  \label{e42}
\end{equation}
at $q=\alpha /2$ (but never converges to the zero value) as $q$ increases up to its allowed maximum value (mandated by $\alpha _{\max }=1$ for each $\ell 
$ value).

\section{Concluding remarks}

In this study, we have shown that a specific transformation/deformation (\ref%
{e02}) of a PGM spacetime (\ref{e01}) effectively yields a von Roos \cite%
{Re271}\ PDM Schr\"{o}dinger equation (\ref{e09.3}). Within such a
deformed/transformed PGM spacetime recipe, we have shown that all our PDM
Schr\"{o}dinger oscillators admit isospectrality and invariance with the
constant mass Schr\"{o}dinger oscillators in the regular PGM spacetime and
in the presence of a Wu-Yang magnetic monopole. Consequently, the exclusive
dependence of the thermodynamical partition function on the energy
eigenvalues manifestly suggests that the Schr\"{o}dinger oscillators and the
PDM Schr\"{o}dinger oscillators share the same thermodynamical properties.
Moreover, we have discussed the hard-wall effects on the energy levels PDM
Schr\"{o}dinger oscillators in a global monopole spacetime without and with
a Wu-Yang magnetic monopole. Drastic energy levels' shift-ups are observed as
a consequence of such hard-wall.

In connection with the energy levels, for both constant mass and PDM Schr\"{o}dinger oscillators in a PGM spacetime and a Wu-Yang magnetic monopole, our observations are in order.  
The common characterization of equal spacing between the energy levels at $\alpha=1$ (flat Minkowski spacetime limit) is only observed for $q=0$ ( no Wu-Yang monopole effect) for all allowed PGM parameter $\alpha$ values (i.e., $0<\alpha \leq 1$). However, for the feasible correlations $q=\alpha /4$, and $q=\alpha /16$ (just two testing toy models), we notice that such equal spacing between energy levels is no longer valid (documented in Figures 1(a), 1(b), and 1(c)). We have also observed that the Wu-Yang monopole yields non-equal spacing between the energy levels (documented in Figures 2(a), 2(b), and 2(c)). Hereby, the energy levels are observed to be shifted up as the PGM parameter $\alpha$ increases for each value of the Wu-Yang monopole parameter $q$ (including $q=0$ for no Wu-Yang monopole). On the other hand, the hard-wall effect is clearly observed through the comparisons between Figures 1(a) and 6(a),  and 1(b) and 6(b). Such comparisons suggest that the equidistance between the energy levels is no longer valid and the separation between the energy levels quadratically increases with increasing PGM parameter $\alpha 
$ (as it increases from just above the zero value). Notably, such drastic shift-ups are expected from the $\alpha ^{2}$- dependence of $%
E_{n_{r},\ell ,q}$ in (\ref{e41}). Nevertheless,  the comparison between Figures 2(a) and 6(c), for a fixed $\alpha =0.5$,  again suggests drastic shift-ups in the energy levels. Moreover,  each energy level slowly converges to the minimum value in (\ref{e42}) at $q=\alpha /2$ (but never converges to the zero value) as $q$ increases up
to its allowed maximum value (mandated by $\alpha _{\max }=1$ for each $\ell 
$ value).

On the thermodynamical properties side, we notice that the Helmholtz free energies 
$f(T)$, Eq.(\ref{e35}), and the mean energy $U\left( T\right) $, Eq. (\ref{e38}), are thermodynamical properties that are directly affected by Wu-Yang \cite{Re101} magnetic monopole, whereas the entropy $S\left( T\right) $, Eq. (\ref{e36}), and the specific heat $%
c\left( T\right) $, Eq. (\ref{e37}), are not.  We have observed that Helmholtz free energies $f(T)$ converge more rapidly to the zero free energy as the Wu-Yang monopole parameter increases with increasing temperature (documented in Figures 3(a), 3(b), and 3(c)). The  mean energy $U\left( T\right) $ decreases for each value of $T$ as the Wu-Yang monopole strength increases through $q=eg$. Yet, we have noticed that the mean energy $U\left( T\right) $, for all allowed $\alpha $ values used, tend to cluster at very high temperatures for $%
q=0$ (i.e., no Wu-yang monopole), and as $q$ increases from zero, such clustering is slowed down (documented in Figures 4(a), 4(b), and 4(c)). On the other hand, the entropy $S\left( T\right) $ increases with increasing temperatures (Figure 5(a)), whereas the specific heat $c\left( T\right) $ increases with increasing temperature up to a maximum value, mandated by the asymptotic behaviour of Eq. (\ref{e37}) so that the ratio $c\left( T\right) /K_{B}\rightarrow 1$ and consequently $c\left( T\right) \rightarrow K_{B}$ as $T\rightarrow \infty$, for all allowed values of the point-like global monopole parameter $\alpha $.  

Finally, the energy levels as well as the thermodynamical properties reported in the current methodical proposal, hold true for both constant mass and PDM\ Schr\"{o}dinger-oscillators in a point-like global monopole spacetime and a Wu-Yang magnetic monopole. This is authorized by the isospectrality and invariance of the two models considered (i.e., constant mass and PDM\ Schr\"{o}dinger-oscillators) in the current study.

\end{document}